\documentclass[11pt,a4paper]{article}
\pdfoutput=1
\usepackage{amsfonts}
\usepackage{lipsum}
\usepackage{amsbsy}
\usepackage{setspace}
\usepackage{blindtext}
\usepackage{bbold}
\usepackage{appendix}
\usepackage{enumerate}
\usepackage{pbox}
\usepackage{cite}
\usepackage{caption}
\usepackage{subcaption}
\usepackage{jheppub}

\title{Does zero temperature decide on the nature of the electroweak phase transition?}

\author[a]{ Christopher~P.~D.~Harman}
\author[a]{ Stephan~J.~Huber}

\affiliation[a]{ Department of Physics and Astronomy, University of Sussex, Brighton, BN1 9QH, UK}

\emailAdd{C.Harman@sussex.ac.uk} 
\emailAdd{S.Huber@sussex.ac.uk}

\abstract{Taking on a new perspective of the electroweak phase transition, we investigate in detail the role played by the depth of the electroweak minimum (``vacuum energy difference''). We find a strong correlation between the vacuum energy difference and the strength of the phase transition. This correlation only breaks down if a negative eigenvalue develops upon thermal corrections in the squared scalar mass matrix in the broken vacuum before the critical temperature. As a result the scalar fields slide across field space toward the symmetric vacuum, often causing a significantly weakened phase transition. Phenomenological constraints are found to strongly disfavour such sliding scalar scenarios. For several popular models, we suggest numerical bounds that guarantee a strong first order electroweak phase transition. The zero temperature phenomenology can then be studied in these parameter regions without the need for any finite temperature calculations. For almost all non-supersymmetric models with phenomenologically viable parameter points, we find a strong phase transition is guaranteed if the vacuum energy difference is greater than $-8.8\times 10^7$~\text{GeV}$^4$. For the GNMSSM, we guarantee a strong phase transition for phenomenologically viable parameter points if the vacuum energy difference is greater than $-6.9\times 10^7$~\text{GeV}$^4$. Alternatively, we capture more of the parameter space exhibiting a strong phase transition if we impose a simultaneous bound on the vacuum energy difference and the singlet mass.}


\arxivnumber{1512.05611}

\begin{document}
\maketitle

\newpage
\pagenumbering{arabic}
\setcounter{page}{1}
\renewcommand{\theequation}{\arabic{section}.\arabic{equation}}

\newpage
\section{Introduction}
Since the discovery of a scalar particle of mass $125~\text{GeV}$ at the Large Hadron Collider \cite{Aad:2012,Chatrchyan:2012}, the question of how electroweak symmetry breaking did happen in the early universe has gained even more urgency. Also the problem remains how to embed the Higgs into a natural framework.

Supersymmetric extensions to the Standard Model (SM) are strong candidates for a fundamental theory that describe observations in particle physics and cosmology \cite{Martin:1997}. These include (a) elegantly unifying all forces at a grand unification scale, (b) providing a dynamical mechanism for electroweak symmetry breaking, and (c) containing a rich dark matter particle sector. Another popular research area in supersymmetric models are the theoretical developments \cite{Pietroni:1992,Davies:1996,Huber:2000,Menon:2004,Huber:2006,Huber:2011,Carena:2008,Huang:2014,Kozaczuk:2014} into obtaining a strong first order electroweak phase transition. Such phase transitions are necessary for electroweak baryogenesis (for a recent review see e.g.~\cite{Morrissey:2012,Konstandin:2013}), i.e. an explanation for the observed matter-antimatter asymmetry of the universe through a mechanism present during the electroweak phase transition.

There is a similar demand for an understanding of how to obtain a strong phase transition in non-supersymmetric models \cite{Grojean:2004,Bodeker:2004,Profumo:2007,Espinosa:2007,Espinosa:2011,Barger:2011,Sannino:2015,Damgaard:2015}. However, there does not currently exist a universal link between a strong phase transition and the zero temperature phenomenology of any given model. One notable work categorises multiple models into three classes, distinguished by whether a strong phase transition is driven by tree level, loop level, or thermal physics \cite{Chung:2012}. A strong phase transition in \cite{Chung:2012} carries the notion of having a large barrier separating the broken and symmetric vacua. They also remark on the zero temperature phenomenology of parameter regions that exhibit a strong phase transition. Our paper adopts a similar approach to studying the electroweak phase transition.

We investigate a new perspective on how to understand the phase transition using a quantity defined at one loop zero temperature: the vacuum energy difference. This very quantity was already mentioned in \cite{Huang:2014}. We investigate in detail the role this quantity plays for some basic properties of the phase transition for six models. These models are described in Section~\ref{section:models} alongside a review of the one loop effective potential at zero temperature and with thermal corrections included. 

Generally, we find a strong correlation between the vacuum energy difference and the strength of the phase transition. This correlation only breaks down if, before the critical temperature, the broken minimum turns into a saddle point upon thermal corrections. This special case can only occur in multi-field models, where it fortunately is further disfavoured once experimental constraints have been applied. So typically a strong first order phase transition is dependent on a mild tuning of the vacuum energy. A tuning at the level of about 30\% is mostly sufficient. This allows one to zoom into the regime of strong first order phase transitions in a simple and efficient way, including for complicated models such as the GNMSSM.

In Section~\ref{section:vac} we define the vacuum energy difference. We then derive analytic expressions of this quantity for all but the supersymmetric model. We discuss the scanning procedure and present the numerical results in Section~\ref{section:num_scan}. The results with and without phenomenological constraints applied are contrasted against each other. Numerical bounds that guarantee a strong phase transition are suggested for phenomenologically viable parameter regions for each model. Three interesting benchmarks scenarios for the GNMSSM data are provided and compared. Finally, we draw up conclusions in Section~\ref{section:conc}.

\section{\label{section:models}The Scalar Potentials}
\subsection{The models}
Throughout this work we will be making reference to the SM, three single field modifications to the SM, and two general singlet extensions of the SM (one of which is supersymmetric). In counting the number of free parameters in each model, we do not include those appearing through one loop corrections from the top and electroweak (EW) gauge bosons\footnote{These SM quantum corrections are governed by the top Yukawa coupling, $y_t$, and EW gauge couplings, $g_2$ and $g_1$.}, each of whose couplings are well determined. We will proceed by briefly describing the models that we use.
\subsubsection*{SM}
For the SM Higgs potential, we use the notation
\begin{equation}
V_{\text{tree}}^{\text{[SM]}}(H)=-\mu_0^2\vert H\vert^2+\lambda_0 \vert H\vert^4\text{,}
\end{equation}
\noindent where $H=(H^+,H^0)$ is the complex SM Higgs doublet and the SM Higgs boson arises from $\phi=\rm{Re}(H^0)$. In setting the Higgs mass to be $m_h=125~\text{GeV}$ and choosing the VEV of $\phi$, we have no free parameters in this model.

\subsubsection*{SM with a dimension-six operator}
We use the potential \cite{Zhang:1992}
\begin{equation}
V_{\text{tree}}^{\text{[SM$+\phi^6$]}}(H)=-\mu_0^2\vert H\vert^2+\lambda_0 \vert H\vert^4+\dfrac{1}{M^2}\vert H\vert^6\text{.}
\end{equation}
\noindent We identify the free parameter of this model as the mass scale, $M$, that appears in the suppression factor of the dimension-six term. The form of this potential can be realised as the low energy description of some strongly coupled models or from integrating out a scalar with a high characteristic mass scale.

\subsubsection*{SM from Gauge Mediation of Exact Scale Breaking (GMESB)}
This model is introduced in ref.~\cite{Abel:2013} as
\begin{equation}
V_{\text{1 loop (0T)}}^{\text{[SM+log]}}(H)=-\dfrac{1}{2}m_h^2\vert H\vert^2\left( 1+\left( \dfrac{4\lambda_0 v^2}{m_h^2}-1 \right)\log\left[ \dfrac{\vert H\vert^2}{v^2} \right] \right)+\lambda_0 \vert H\vert^4\text{.}
\end{equation}
\noindent This potential is the quantum effective potential at zero temperature. It arises when the scale symmetry is broken in a hidden sector through quantum corrections and mediated to the observable sector via gauge interactions only. We identify the free parameter of this theory to be the quartic self-coupling of the Higgs, $\lambda_0$. The phase transition of this model has previously been studied in \cite{Dorsch:2014}.

\subsubsection*{SM with an additional Coleman-Weinberg scalar}
We use the same potential as that of the SM but include a new scalar that contributes a Coleman-Weinberg term at zero temperature
\begin{equation}
V_{\text{1 loop (0T)}}^{\text{[SM+scalar]}}(H)=V_{\text{1 loop (0T)}}^{\text{[SM]}}(H)+\dfrac{1}{(8\pi)^2}m_X^4(H)\left( \log\left[ \dfrac{m_X^2(H)}{Q^2}\right]-\dfrac{3}{2} \right)\text{,}
\end{equation}
\noindent where $m_X^2(H)=y^2 \vert H\vert^2$. The $V_{\text{1 loop (0T)}}^{\text{[SM]}}(H)$ term is the SM one loop Higgs potential. The free parameter of this theory is the coupling, $y$, of the new scalar to the Higgs. We make the additional, somewhat artificial assumption that the new scalar does not produce thermal corrections to the potential. We use this model as a probe to distinguish between the impact of zero and finite temperature corrections to the effective potential.

\subsubsection*{SM plus a real singlet (xSM)}
We write the potential with a similar notation to ref.~\cite{Profumo:2007}
\begin{equation}
V_{\text{tree}}^{\text{[xSM]}}(H,S)=-\mu_0^2\vert H \vert^2+\lambda_0 \vert H \vert^4+\dfrac{a_1}{2}\vert H \vert^2S+\dfrac{a_2}{2}\vert H \vert^2S^2+\dfrac{b_2}{2}S^2+\dfrac{b_3}{3}S^3+\dfrac{b_4}{4}S^4\text{.}
\end{equation}
\noindent Here $S$ is a real singlet scalar field. This potential contains three types of terms: purely $H$, purely $S$, and mixed terms. Note that we have cubic terms entering as both an $S^3$ and $S\vert H \vert^2$ term. Essential to phenomenological constraints is the Higgs-singlet mixing angle, $\sin\alpha$, defined via
\begin{equation}
\left(\begin{matrix}
h\\
s
\end{matrix}\right)=
\left(\begin{matrix}
\cos\alpha & \sin\alpha\\
-\sin\alpha & \cos\alpha
\end{matrix}\right)
\left(\begin{matrix}
\phi\\
S
\end{matrix}\right)
\text{.}
\label{shmix}
\end{equation}
\noindent We can recognise $\sin\alpha$ as the singlet component of the $h$-state. In rewriting the parameters $\mu_0$, $a_2$, $b_2$, and $b_4$ in terms of $v$, $v_S$, $m_{h}$, and $m_{s}$ (of which $v$ and $m_h$ are fixed) we are left with a total of five free parameters (two of them being tree level cubic terms). We will define the new parameter choice more precisely in Section~\ref{section:xSMvac}.

\subsubsection*{GNMSSM}
Supersymmetric extensions of the SM are promising settings to realise a strong phase transition. However, in the Minimal Supersymmetric extension to the Standard Model (MSSM) with superpotential \cite{Martin:1997}
\begin{equation}
\mathcal{W}_{\text{MSSM}} = \bar{u}\mathbf{y_u}QH_u-\bar{d}\mathbf{y_d}QH_d-\bar{e}\mathbf{y_e}LH_d+\mu H_u \cdot H_d,
\end{equation}
LHC constraints on Higgs properties make a strong phase transition driven by light stops very unlikely \cite{Carena:2012}. Here $\bar{u},~\bar{d},~\bar{e},~Q$ and $L$ are the usual lepton and quark supermultiplets, $\mathbf{y_u},~\mathbf{y_d}$, and $\mathbf{y_e}$ are $3\times 3$ Yukawa matrices, $H_u=( H_u^+,H_u^0)$ and $H_d=( H_d^0,H_d^-)$ are the ``up-type'' and ``down-type'' complex Higgs doublets, and $\mu$ is the supersymmetric analogue of the Higgs mass, commonly referred to as the ``$\mu$-parameter''.

Singlet extensions of the MSSM have attractive features for Higgs phenomenology. For instance, there are mechanisms to increase the natural upper bound of the lightest CP-even Higgs bosons mass (see e.g. \cite{Ellwanger:2009,Ross:2012}). Also these models often generate a strong phase transition \cite{Kozaczuk:2014,Huang:2014,Pietroni:1992,Davies:1996,Huber:2000,Menon:2004,Huber:2006ma}. Singlet extensions of the MSSM are often distinguished by discrete symmetries. Here we study the most general singlet extension, the Generalised Next-to-Minimal Supersymmetric extension to the Standard Model (GNMSSM) with the superpotential
\begin{equation}
\mathcal{W}=\mathcal{W}_{\text{MSSM}}+\lambda S H_u \cdot H_d+k_1S+\frac{1}{2}k_2S^2+\frac{1}{3}k_3 S^3\text{,}
\end{equation}
\noindent where $S$ is a chiral singlet superfield and $\lambda$, $k_1$, $k_2$, and $k_3$ encode couplings and masses. This model can be derived in a top-down approach based on a discrete R symmetry as shown in \cite{Lee:2011}. Not having a discrete symmetry automatically evades a possible domain wall problem that plagues more constrained setups \cite{Abel:1995}. Adding the usual soft supersymmetry breaking terms, the tree level scalar potential is given by
\small
\begin{equation}
\begin{array}{l l}
V_{\text{tree}}^{\text{[GNMSSM]}}= & \displaystyle \left( \vert \mu+\lambda S \vert^2 +m_{H_u}^2\right) \vert H_u \vert^2+\left( \vert \mu+\lambda S \vert^2 +m_{H_d}^2\right) \vert H_d \vert^2+m_S^2\vert S \vert^2\\
 & \\
 & \displaystyle+\vert \lambda H_u \cdot H_d+k_1+k_2S+k_3S^{2} \vert^2+\frac{1}{8}(g_2^2+g_1^2)\left( \vert H_u \vert^2-\vert H_d \vert^2 \right)^2+\frac{1}{2}g_2^2\vert H_d^\dagger H_u \vert^2 \\
 & \\
& \displaystyle+\left[\left( (b\mu + A_\lambda \lambda S) H_u \cdot H_d+A_{k_1} k_1 S+\frac{1}{2}A_{k_2}k_2S^2+\frac{1}{3}A_{k_3}k_3S^3 \right)+h.c\right]\text{.}
\end{array}
\label{GNMSSM_tree_pot}
\end{equation}
\normalsize
\noindent We decompose the Higgs gauge-eigenstates into the mass-eigenstates via
\begin{equation}
\left(\begin{matrix}
H_u^0\\
H_d^0\\
S
\end{matrix}\right)
=
\left(\begin{matrix}
 v\sin\beta \\
 v\cos\beta \\
 v_S
\end{matrix}\right)
+\dfrac{1}{\sqrt{2}}
\mathbf{R}_{0^+}
\left(\begin{matrix}
h^0\\
H^0\\
s^0
\end{matrix}\right)
+\dfrac{i}{\sqrt{2}}
\mathbf{R}_{0^-}
\left(\begin{matrix}
G^0\\
A^0\\
\eta^0
\end{matrix}\right),
\label{neutGaugeMassGNMSSM}
\end{equation}
\begin{equation}
\left(\begin{matrix}
H_u^+\\
H_d^{-*}
\end{matrix}\right)
=
\mathbf{R}_{\pm}
\left(\begin{matrix}
G^+\\
H^+
\end{matrix}\right)\text{,}
\end{equation}
\noindent where the $\mathbf{R}$'s are the relevant rotation matrices. We only need to understand the contents of $\mathbf{R}_{0^+}$ (the CP-even mass mixing matrix) for this study. In order to more easily compare the phenomenology of the GNMSSM to that of the xSM, we decouple the heavy CP-even Higgs boson, $H^0$. In practice this means that we reduce the three-dimensional field space of eq.~(\ref{neutGaugeMassGNMSSM}) into a two-dimensional field space by looking in the $\tan\beta$ direction
\begin{equation}
\left(\begin{matrix}
H_u^0\\
H_d^0\\
S
\end{matrix}\right)=
\left(\begin{matrix}
\phi \sin\beta\\
\phi \cos\beta\\
S
\end{matrix}\right)=
\left(\begin{matrix}
\sin\beta & 0\\
\cos\beta & 0\\
0 & 1
\end{matrix}\right)
\left(\begin{matrix}
\phi\\
S
\end{matrix}\right)
\text{.}
\label{tanbetaDir}
\end{equation}
\noindent We will be scanning for parameter points where $\tan\beta$ varies from low to medium values so we will keep the $\beta$-dependence explicit throughout this work. Just as in the xSM, we recognise $\sin\alpha$ as the singlet component of the CP-even Higgs state, $h^0$. We allow for either the lightest or next-to-lightest state \cite{Ellwanger:2009} to be $h^0$, recognised as the 125~\text{GeV} Higgs boson.

It is well known that Higgs sectors of supersymmetric extensions to the SM suffer from a tree level bound on the lightest CP even state (see ref.~\cite{Martin:1997} for a review). Radiative corrections from the stop sector are crucial. The stop squared-masses are given by
\begin{equation}
\begin{array}{r l}
m_{\tilde{t}_1}^2=&\displaystyle \dfrac{1}{2} \left(m_{\tilde{t}_L}^2+m_{\tilde{t}_R}^2 + \sqrt{(m_{\tilde{t}_L}^2-m_{\tilde{t}_R}^2)^2+4 m_{X_t}^4} \right) \\
 & \\
m_{\tilde{t}_2}^2=&\displaystyle\dfrac{1}{2} \left(m_{\tilde{t}_L}^2+m_{\tilde{t}_R}^2 - \sqrt{(m_{\tilde{t}_L}^2-m_{\tilde{t}_R}^2)^2+4 m_{X_t}^4} \right)\text{,}
\end{array}
\end{equation}
\noindent where the squared-mass matrix in the gauge-eigenstate basis $(\tilde{t}_L,\tilde{t}_R)$ is given by
\small
\begin{equation}
\mathbf{m_{\tilde{t}}^2}=
\left( \begin{array}{c c}
\displaystyle m_{\tilde{t}_L}^2=m_{Q_3}^2+y_t^2 \vert H_u^0\vert^2+\Delta_{\tilde{u}_L} & \displaystyle m_{X_t}^2=A_t^*y_t (H_u^0)^* - (\mu+\lambda S) y_t H_d^0\\
& \\
\displaystyle (m_{X_t}^2)^*=A_ty_t H_u^0 - (\mu+\lambda S)^* y_t (H_d^0)^* & \displaystyle m_{\tilde{t}_R}^2=m_{\bar{u}_3}^2+y_t^2 \vert H_u^0\vert^2+\Delta_{\tilde{u}_R}
\end{array} \right)\text{,}
\label{stopGauge}
\end{equation}
\normalsize
\begin{equation}
\text{and }
\Delta_{\tilde{u}_L}=\dfrac{1}{4}\left(g_2^2-\frac{1}{3}g_1^2\right)\left(\vert H_u^0 \vert^2-\vert H_d^0 \vert^2\right)
\text{, }
\Delta_{\tilde{u}_R}=\dfrac{1}{3}g_1^2\left(\vert H_u^0 \vert^2-\vert H_d^0 \vert^2\right)\text{,}
\end{equation}

\noindent and $m_{Q_3}$ and $m_{\bar{u}_3}$ are the stop soft masses, $A_t$ is a third generation soft parameter, and $y_t$ is the top Yukawa coupling.

Assuming there are no CP violating phases and all terms in the potential are real, we have a total of 16 parameters in this theory. However, not all of these are free parameters. Applying the minimum conditions and assuming that only the real parts of the fields are non-zero in the minimum, we reparameterise the Higgs mass-squared soft parameters ($m_{H_u}^2$, $m_{H_d}^2$, and $m_{S}^2$) by the VEVs in the broken phase ($v$, $\tan\beta$, and $v_S$). We also choose to remove the singlet linear term in the potential by taking $A_{k_1}=-k_2$, meaning that a local extremum will exist at the zero point in field space. Finally, we choose a special setup for the stop soft parameters. Namely that we fix $A_t=(\mu+\lambda v_S)\cot\beta$ so that the off-diagonal elements of eq.~(\ref{stopGauge}) vanish at the broken minimum. Furthermore, we impose that the stop soft mass parameters are nearly degenerate, $m_{Q_3}-m_{\overline{u}_3}=100~\text{GeV}$. The value of $m_{Q_3}$ is fixed such that we have a suitable Higgs with mass $125~\text{GeV}$. We then count a total of 11 free parameters in this theory. A brief summary of the scan procedure can be found in Appendix~\ref{section:paraGNMSSM}.

\subsection{At one loop zero temperature}
The general form of the one loop zero temperature effective potential in the models we study is
\begin{equation}
\begin{array}{l l}
V_{\text{1 loop (0T)}}(\phi,S)& = V_{\text{tree}}(\phi,S)+V_{\text{CT}}(\phi,S)+V_{\text{CW}}(\phi,S)\text{,}\\
\end{array}
\label{oneLoopEffPot}
\end{equation}
\noindent where $\phi=\rm{Re}(H^0)$ is the SM-like Higgs field and $S$ is a singlet field under each of the SM gauge groups. The individual terms are given by
\begin{equation}
\begin{array}{r l}
V_{\text{tree}}(\phi,S) &= \displaystyle -\mu_0^2\phi^2+\lambda_0 \phi^4+V_{\text{tree}}^{\text{[non-SM]}}(\phi,S)\text{,}\\
 & \\
V_{\text{CT}}(\phi,S) &= \displaystyle \dfrac{1}{2}\delta m_\phi^2 \phi^2+\dfrac{1}{2}\delta m_S^2 S^2+\dfrac{1}{4}\delta \lambda_0 \phi^4\text{,}\\
 & \\
V_{\text{CW}}(\phi,S) &= \displaystyle \dfrac{1}{(8\pi)^2}\sum_i g_i(-1)^{2s_i} m_i^4(\phi,S)\left[ \log\left( \dfrac{m_i^2(\phi,S)}{Q^2} \right)-\dfrac{3}{2} \right]\text{,}
\end{array}
\end{equation}
\noindent where the $\delta$'s are the one loop counter terms and the index $i$ runs over all bosons and fermions, with $g_i$ degrees of freedom and spin $s_i$, considered at one loop. Note that we use the Coleman-Weinberg (CW) effective potential in the modified $\overline{\text{DR}}$ scheme \cite{Jack:1994}, and $Q$ is the renormalisation scale, chosen to be the mass of the top quark, $m_t$, throughout this investigation. Unless otherwise stated, we will adopt the convention that the VEVs of $\phi$ and $S$ at zero temperature are given by $\langle \phi \rangle=v=174.2~\text{GeV}$ and $\langle S \rangle=v_S$, respectively, and denote the pole mass of the $i^{th}$ particle by $m_i = m_i(\phi=v,S=v_S)$.

\noindent We choose the renormalisation conditions
\begin{equation}
\left.\dfrac{\partial V_{\text{tree}}}{\partial \phi}\right\vert_{\text{broken}}=\left.\dfrac{\partial V_{\text{1 loop (0T)}}}{\partial \phi}\right\vert_{\text{broken}}\text{,}
\end{equation}
\begin{equation}
\left.\dfrac{\partial V_{\text{tree}}}{\partial S}\right\vert_{\text{broken}}=\left.\dfrac{\partial V_{\text{1 loop (0T)}}}{\partial S}\right\vert_{\text{broken}}\text{,}
\end{equation}
\begin{equation}
\left.\dfrac{\partial^2 V_{\text{tree}}}{\partial \phi^2}\right\vert_{\text{broken}}=\left.\dfrac{\partial^2 V_{\text{1 loop (0T)}}}{\partial \phi^2}\right\vert_{\text{broken}}\text{.}
\label{higgrenorm}
\end{equation}
\noindent The condition in eq.~(\ref{higgrenorm}) means that the Higgs mass is unchanged upon radiative corrections. This condition cannot be applied for the GNMSSM due to the tree level bound on the lightest CP-even Higgs state. Therefore $\delta \lambda_0=0$ for the GNMSSM so the Higgs mass is left to run. The other two conditions keep the VEVs in the broken minimum the same at one loop as their tree level values.
{
Note that we have chosen renormalisation conditions only in the broken phase, which is sufficient for our purpose. For a more general analysis, including renormalisation conditions related to the symmetric phase, see ref.~\cite{Huber:2015}.}

\subsection{At one loop finite temperature}
In order to study cosmological phase transitions in a quantum field theory framework, the one loop effective potential ought to take into account a temperature-dependent piece. We include thermal corrections at one loop, such that the thermal effective potential reads
\begin{equation}
V_{\text{1 loop}}(\phi,S;T) = V_{\text{1 loop (0T)}}(\phi,S)+V_{\text{T}}(\phi,S;T)\text{,}
\end{equation}
\noindent where \cite{Dolan:1973}
\small
\begin{equation}
V_{\text{T}}(\phi,S;T)=\sum_{i=f,b}^\infty \dfrac{(-1)^{2s_i}g_i T^4}{2\pi^2}\int_0^\infty dx x^2 \log \left[1+(-1)^{2s_i+1} \exp\left( -\sqrt{x^2+\dfrac{m_i^2(\phi,S)}{T^2}} \right) \right]\text{,}
\label{exact_therm_loop}
\end{equation}
\normalsize
\noindent and $T$ is the temperature of the surrounding plasma. The sum is over all relevant fermions and bosons in the plasma. Rather than numerically evaluating the integral in eq.~(\ref{exact_therm_loop}), we will use the potential in the form of a piecewise function built up of three parts as described below. Each part is determined by the value of $m_i(\phi,S)/T$ for each particle. Note that we are going to mostly focus on the limit of very strong phase transitions, where thermal resummations \cite{Arnold:1992} of the potential do not play a crucial role. So we ignore these.

The potential in eq.~(\ref{exact_therm_loop}) can be rewritten into an analytic form within two approximations: a low temperature limit, where $m_i(\phi,S)/T$ is large, and a high temperature limit, where $m_i(\phi,S)/T$ is small \cite{Dolan:1973,Anderson:1991}. We use interpolation functions for intermediate temperatures, during which the low and high temperature approximations differ from the exact value by no more than $4\%$. The analytic form of these finite temperature contributions depends on whether the $i^{th}$ particle is a boson or a fermion. Notably, only bosonic thermal contributions contain temperature-dependent cubic terms which may alter the strength of the phase transition. All field-dependence appears through the field-dependent mass of the contributing particle, $m_i(\phi,S)$.

For early universe considerations, such as electroweak baryogenesis, we are interested in the strength of the phase transition. In this work, the critical temperature is defined at when the electroweak symmetry broken and symmetric vacua are degenerate. Given the chosen VEV convention, a strong phase transition is defined\footnote{The factor of $\sqrt{2}$ accounts for the chosen normalisation of the Higgs field. This condition satisfies the baryon preservation criteria \cite{Kuzmin:1985,Moore:1998}.} by $\sqrt{2}v_c/T_c>1$. Here $v_c$ is the value of the $\phi$ field in the broken vacuum at critical temperature $T_c$. Henceforth we shall denote the strength of the phase transition by the shorthand, $\xi=\sqrt{2}v_c/T_c$, for convenience.

\section{\label{section:vac}The Vacuum Energy Difference}
We define the one loop vacuum energy difference at zero temperature between the broken and symmetric vacuum to be
\begin{equation}
\begin{array}{l l}
\Delta V_{\text{1 loop (0T)}}& \displaystyle=\left.V_{\text{1 loop (0T)}}\right\vert_{\text{broken}}-\left.V_{\text{1 loop (0T)}}\right\vert_{\text{symmetric}}\\
 & \\
 & \displaystyle= V_{\text{1 loop (0T)}} (v,v_S) -V_{\text{1 loop (0T)}}(0,\tilde{v}_S)\\
 & \\
 & \displaystyle= \Delta V_{\text{tree}}+\Delta V_{\text{rad}}\text{,}
\end{array}
\label{vac_en_diff}
\end{equation}
\noindent where we have defined the quantities
\begin{equation}
\begin{array}{r l}
\Delta V_{\text{tree}} =& \displaystyle V_{\text{tree}}\left(v,v_S\right)-V_{\text{tree}}\left(0,\tilde{v}_S\right)\text{,}\\
 & \\
\Delta V_{\text{rad}} =& \displaystyle \left[V_{\text{CT}}+\Delta V_{\text{CW}}\right]\left(v,v_S\right)-\left[V_{\text{CT}}+\Delta V_{\text{CW}}\right]\left(0,\tilde{v}_S \right)\text{,}
\end{array}
\end{equation}
\noindent and $\tilde{v}_S$ is the value of the singlet field $S$ in the symmetric vacuum. Note that the vacuum energy difference takes on negative values if the broken minimum is the global minimum of the potential.

The potential difference between the symmetric and broken minima is temperature-dependent. The critical temperature is defined as the temperature at which this potential difference is zero. The suggestion we want to investigate in the following work is therefore:
\begin{center}
\textit{The smaller the value of }$\vert \Delta V_{\text{1 loop (0T)}}\vert$\textit{, the stronger the phase transition.}
\end{center}
\noindent A decrease in $\vert\Delta V_{\text{1 loop (0T)}}\vert$ is expected to decrease the critical temperature and therefore increases the strength of the phase transition $\xi$. The concept of the vacuum energy difference is a more precise prescription of the notion of ``flat potentials'' in ref.~\cite{Sannino:2015}.

As we will see below, the one loop vacuum energy difference is often simply related to the free parameters of the models we investigate. In each model, we consider one loop (zero temperature and thermal) contributions from the top quark, $t$, and the EW gauge bosons, $W^\pm$ and $Z^0$. In the GNMSSM, we also consider the one loop (zero temperature and thermal) contributions from the stops, $\tilde{t}_1$ and $\tilde{t}_2$, the supersymmetric partners of the SM top quark.

{
In this work we approximate the effective potential at one loop. The impact of higher loop orders on the effective potential is model dependent. We expect higher loop order corrections to be more relevant for Coleman-Weinberg type models, where radiative corrections play a large role in determining the shape of the potential. This is not the case for most of the models we explore since the depth of the broken vacuum is effectively set by the observed Higgs mass and well approximated by the tree level potential. A possible exception is the GNMSSM, where the Higgs mass receives crucial one loop contributions. 

Note that we should remain cautious regarding the gauge-dependence of our results \cite{Patel:2011,Wainwright:2011,Wainwright:2012}. Interesting enough, reference \cite{Andreassen:2014} suggests that for certain models the potential evaluated at its true minimum is gauge-{\em invariant} at one-loop. Such works ought to be taken further to quantify whether this is true for each model we explore.}

\subsection{The vacuum energy difference in single field models}
We apply the minimum condition and use the Higgs mass to rewrite the quartic coupling. In the SM, this means $\mu_0^2=2\lambda_0v^2$ and $m_h^2=4\lambda_0v^2$. We then read off the tree level vacuum energy difference as
\begin{equation}
\Delta V_{\text{tree}}^{\text{[SM]}}=-\lambda_0 v^4=-\dfrac{1}{4}m_h^2 v^2\text{.}
\end{equation}
Including the top quark, $W^\pm$, and $Z^0$-boson one loop corrections, we find the one loop zero temperature vacuum energy difference to be
\begin{equation}
\begin{array}{l l}
\Delta V_{\text{1 loop (0T)}}^{\text{[SM]}} & = -\dfrac{1}{4}m_h^2 v^2-\dfrac{2}{(16\pi)^2} \left( g_tm_t^4-g_Wm_W^4-g_Zm_Z^4 \right)\\
 & \\
 & = -1.185 \times 10^8~\text{GeV}^4-\dfrac{2}{(16\pi)^2} \left( g_tm_t^4-g_Wm_W^4-g_Zm_Z^4 \right)\\
 & \\
 & \displaystyle = -1.267 \times 10^8~\text{GeV}^4\text{.}
\end{array}
\label{SMvac}
\end{equation}

\noindent We see that quantum corrections do not drastically affect the vacuum energy difference in the SM. The top quark dominates the radiative correction and decreases the vacuum energy difference by 7.2\%. Including the EW gauge bosons, it decreases by 6.9\%. In other words, the vacuum energy difference in the SM is effectively set by the Higgs mass (the tree level contribution).

Let us repeat this procedure for other extensions of the SM. For the SM with a dimension-six term
\begin{equation}
\Delta V_{\text{1 loop (0T)}}^{\text{[SM$+\phi^6$]}} = -\dfrac{1}{4}m_h^2 v^2+\dfrac{v^6}{M^2}-\dfrac{2}{(16\pi)^2} \left( g_tm_t^4-g_Wm_W^4-g_Zm_Z^4 \right)\text{,}
\end{equation}
\noindent for the SM from GMESB
\begin{equation}
\Delta V_{\text{1 loop (0T)}}^{\text{[SM+log]}} = -\dfrac{1}{2}m_h^2 v^2+\lambda_0v^4\text{,}
\end{equation}
\noindent and for the SM with an additional CW scalar
\begin{equation}
\Delta V_{\text{1 loop (0T)}}^{\text{[SM+scalar]}} = -\dfrac{1}{4}m_h^2 v^2-\dfrac{2}{(16\pi)^2} \left( g_tm_t^4-g_Wm_W^4-g_Zm_Z^4-y^4v^4 \right)\text{.}
\end{equation}
\noindent In all these models, the vacuum energy difference can be chosen independently of the Higgs mass.

\subsection{\label{section:xSMvac}The vacuum energy difference (xSM)}
\noindent Applying the minimum conditions,

$$
\left.\dfrac{\partial V_{\text{tree}}^{\text{[xSM]}}}{\partial \phi}\right\vert_{\phi\rightarrow v,S\rightarrow v_S}=0~~~ \text{ and}~~~
\left.\dfrac{\partial V_{\text{tree}}^{\text{[xSM]}}}{\partial S}\right\vert_{\phi\rightarrow v,S\rightarrow v_S}=0 \text{,}
$$
\noindent we find
$$
\mu^2=2\lambda_0v^2-\left(\dfrac{a_1}{2v_S}+\dfrac{a_2}{2}\right)v_S^2\text{ and }
b_2=-\left( \dfrac{a_1}{2v_S}+a_2 \right)v^2-\left( \dfrac{b_3}{v_S}+b_4 \right)v_S^2\text{.}
$$

\noindent This gives us a tree level vacuum energy difference of
\begin{equation}
\begin{array}{r l}
\Delta V_{\text{tree}}^{\text{[xSM]}} =& -\lambda_0v^4-\left[ \left( \dfrac{a_1}{2v_S}+a_2 \right)v^2+\dfrac{b_4}{2}\left( v_S^2-\tilde{v}_S^2 \right) \right]\left(\dfrac{v_S^2-\tilde{v}_S^2}{2}\right)\\
 & \\
 & -\dfrac{b_3}{6}\left( v_S-\tilde{v}_S \right)^2\left( v_S+2\tilde{v}_S \right)\text{,}
\end{array}
\end{equation}

\noindent where the singlet VEV in the symmetric vacuum is given by
\begin{equation}
\tilde{v}_S=-\dfrac{b_3}{2b_4}\pm\sqrt{\left( \dfrac{b_3}{2b_4} \right)^2-\dfrac{b_2}{b_4}} \text{.}
\label{Sext}
\end{equation}

\noindent The sign in eq.~(\ref{Sext}) is determined by whichever minimum has the lowest value of the potential. All one loop contributions considered here are the same as those in the SM. The one loop zero temperature vacuum energy difference is therefore given by
\begin{equation}
\begin{array}{r l}
\Delta V_{\text{1 loop (0T)}}^{\text{[xSM]}} = & -\lambda_0 v^4-\dfrac{2}{(16\pi)^2} \left( g_tm_t^4-g_Wm_W^4-g_Zm_Z^4 \right)\\
 & \\
 & -\left[ \left( \dfrac{a_1}{2v_S}+a_2 \right)v^2+\dfrac{b_4}{2}\left( v_S^2-\tilde{v}_S^2 \right) \right]\left(\dfrac{v_S^2-\tilde{v}_S^2}{2}\right)\\
 & \\
 & -\dfrac{b_3}{6}\left( v_S-\tilde{v}_S \right)^2\left( v_S+2\tilde{v}_S \right)\text{.}
\end{array}
\label{xSMvac1}
\end{equation}

\noindent The first line of eq.~(\ref{xSMvac1}) is algebraically identical to the SM vacuum energy difference at one loop prior to fixing $\lambda_0$ in favour of the SM Higgs mass, $m_h$. Note that in the case of $\tilde{v}_S=v_S$, we recover the SM result.

We rewrite the quartic terms, $a_2$ and $b_4$, in favour of the CP even mass eigenstates, $m_{\phi_1}$ and $m_{\phi_2}$, where $m_{\phi_1} < m_{\phi_2}$. Both $m_{\phi_1}$ and $m_{\phi_2}$ are recognised with the SM-like Higgs mass and singlet mass ($m_h$ and $m_s$, respectively) depending on the ordering of their mass values. Therefore
\begin{equation}
(a_2)_\pm=-\dfrac{a_1}{2v_S} \pm \dfrac{1}{v_Sv}\sqrt{\left( m_{\phi_1}^2-4\lambda_0v^2 \right)\left( 4\lambda_0v^2-m_{\phi_2}^2 \right)}\text{ and}
\label{a1xSM}
\end{equation}
\begin{equation}
b_4=\dfrac{1}{v_S^2}\left[ m_{\phi_1}^2+m_{\phi_2}^2-4\lambda_0v^2+\dfrac{a_1}{2}\dfrac{v^2}{v_S}-\dfrac{b_3}{3}v_S \right]\text{.}
\end{equation}
\noindent Given that the quartic coupling, $a_2$, must be a real-valued quantity, we find
\begin{equation}
m_{\phi_1}^2 \leq 4\lambda_0 v^2 \leq m_{\phi_2}^2\text{.}
\label{lambda_ineq}
\end{equation}
\noindent Altogether, we find the one loop zero temperature vacuum energy difference to be
\begin{equation}
\begin{array}{r l}
\Delta V_{\text{1 loop (0T)}}^{\text{[xSM]}} = & \displaystyle -\lambda_0v^4-\left[ \pm \dfrac{v}{v_S} \sqrt{(m_h^2-4\lambda_0v^2)(4\lambda_0v^2-m_s^2)}\right.\\
 & \\
 & \left.+\dfrac{1}{2}\left( m_h^2+m_s^2-4\lambda_0 v^2+\dfrac{a_1 v^2}{2v_S}-\dfrac{b_3}{3}v_S \right)\left( 1-\dfrac{\tilde{v}_S^2}{v_S^2} \right) \right]\left(\dfrac{v_S^2-\tilde{v}_S^2}{2}\right)\\
 & \\
 & \displaystyle  -\dfrac{b_3}{6}\left( v_S-\tilde{v}_S \right)^2\left( 2\tilde{v}_S+v_S \right)-\dfrac{2}{(16\pi)^2} \left( g_tm_t^4-g_Wm_W^4-g_Zm_Z^4 \right)\text{.}
\end{array}
\label{xSMvac2}
\end{equation}
Identifying the free parameters, the above expression contains the two cubic terms ($a_1$ and $b_3$), two physical Higgs masses ($m_{h}$ and $m_{s}$), three VEVs ($v$, $v_S$ and $\tilde{v}_S$), and the quartic Higgs self-coupling ($\lambda_0$). We can again see that we are free to choose the vacuum energy difference, via the free parameters of the model, despite the Higgs mass being fixed.

\subsubsection*{$\mathbb{Z}_2$ symmetric case (with broken $\mathbb{Z}_2$ at zero temperature)}
\noindent By imposing a $\mathbb{Z}_2$ discrete symmetry on the singlet, the cubic terms vanish, giving a model referred to as the $\mathbb{Z}_2$xSM. Setting the cubic terms to zero in eq.~(\ref{xSMvac2}), we find a simple expression for the one loop vacuum energy difference at zero temperature,
\small
\begin{equation}
\Delta V_{\text{1 loop (0T)}}^{\text{[$\mathbb{Z}_2$xSM]}}=-\dfrac{1}{4}m_h^2v^2\left( 1+\dfrac{m_h^2-4\lambda_0v^2}{m_s^2} \right)^{-1}-\dfrac{2}{(16\pi)^2}\left( g_tm_t^4-g_Wm_W^4-g_Zm_Z^4 \right)\text{.}
\label{Z2xSm_raw1}
\end{equation}
\normalsize
\noindent Note that this expression assumes that $v_S$ is non-zero, so the $\mathbb{Z}_2$ symmetry is spontaneously broken. This expression is almost identical to the SM expression in eq.~(\ref{SMvac}) with the exception of a multiplicative factor on the tree level term. For this factor to be less than one we must have $4\lambda_0 v^2\leq m_h^2$, hence $m_s<m_h$ is the only way in which a vacuum energy difference higher than the SM can be obtained. A strange feature is that eq.~(\ref{Z2xSm_raw1}) is independent of the potential's structure in the singlet direction: only $m_s$ and $\lambda_0$ appear as free parameters in the vacuum energy difference.

Let us replace $\lambda_0$ by a new parameter, $\epsilon$, defined by
\begin{equation}
4\lambda_0 v^2=\epsilon m_h^2+(1-\epsilon)m_s^2 \text{.}
\label{defeps}
\end{equation}
\noindent The inequality of eq.~(\ref{lambda_ineq}) translates into $0\leq \epsilon \leq 1$. This allows us to rewrite the vacuum energy difference in the $\mathbb{Z}_2$xSM model as
\begin{equation}
\Delta V_{\text{1 loop (0T)}}^{\text{[$\mathbb{Z}_2$xSM]}} = -\dfrac{1}{4}\left[\dfrac{m_h^2m_s^2}{(1-\epsilon)m_h^2+\epsilon m_s^2} \right]v^2-\dfrac{2}{(16\pi)^2}\left( g_tm_t^4-g_Wm_W^4-g_Zm_Z^4 \right)\text{.}
\label{Z2xSM vac energy diff}
\end{equation}

\noindent The lowest value for $\vert\Delta V_{\text{1 loop (0T)}}^{\text{[$\mathbb{Z}_2$xSM]}}\vert$ is bounded by the one loop contribution. This happens when the tree level contribution vanishes, which is only possible if $m_s$ goes to zero. Furthermore, we can rewrite the tree level potential such that the importance of $\epsilon$ is clearer,
\begin{equation}
\begin{array}{r l}
V_{\text{tree}}^{\text{[$\mathbb{Z}_2$xSM]}}= & \displaystyle \dfrac{1}{2}m_h^2\left[ \left( \dfrac{\phi^2}{2v^2}-1 \right)\phi^2\epsilon+\left( \dfrac{S^2}{2v_S^2}-1 \right)S^2(1-\epsilon) \right]\\
 & \\
 & \displaystyle +\dfrac{1}{2}m_s^2\left[ \left( \dfrac{S^2}{2v_S^2}-1 \right)S^2\epsilon+\left( \dfrac{\phi^2}{2v^2}-1 \right)\phi^2(1-\epsilon) \right]\\
 & \\
 & \displaystyle \pm \dfrac{1}{2}(m_h^2-m_s^2)\sqrt{\epsilon(1-\epsilon)}\left[ \dfrac{v_S}{v}\phi^2+\dfrac{v}{v_S}S^2-\dfrac{1}{v_Sv}\phi^2S^2 \right]\text{.}
\end{array}
\label{Z2xSM_pot}
\end{equation}
\noindent In the limit that $\epsilon$ goes to unity (zero), the tree level potential collapses to that of the SM in the $\phi$ (singlet) direction. The other piece of the potential corresponds to an invisible sector that is phenomenologically inaccessible since the $\phi$ and $S$ fields no longer mix. Thus we expect the $\mathbb{Z}_2$xSM to behave in a similar manner to the SM close to these limits. Taking the tree level piece of eq.~(\ref{Z2xSM vac energy diff}) and solving for the singlets mass, we find
\begin{equation}
m_s=m_h\left(1+\dfrac{1}{1-\epsilon}\left[ \dfrac{\Delta V_{\text{tree}}^{\text{[SM]}}}{\Delta V_{\text{tree}}^{[\text{$\mathbb{Z}_2$xSM]}}} -1 \right]\right)^{-1/2}.
\label{mSZ2}
\end{equation}
\noindent If we take $\Delta V_{\text{tree}}^{[\text{$\mathbb{Z}_2$xSM]}} \rightarrow 0$, then eq.~(\ref{mSZ2}) suggests that the singlets mass vanishes irrespective of the value of $\epsilon$. For the case of $\epsilon=0$, the singlet mass is determined by the vacuum energy difference, since $V_{\text{tree}}^{[\text{$\mathbb{Z}_2$xSM]}}=-m_s^2v^2/4$. For the case of $\epsilon=1$, it naively appears that the singlet mass must be zero and we recover the SM. However, there is one special parameter choice that allows the SM Higgs and singlet fields to coexist. This happens if $v_S=0$, whereby the two fields decouple yet the mixing term does not disappear. The limit $\epsilon\rightarrow1$ in eq.~(\ref{mSZ2}) is no longer so trivial.

\subsubsection*{$\mathbb{Z}_2$ symmetric case (with unbroken $\mathbb{Z}_2$ at zero temperature)}
In the special case of a $\mathbb{Z}_2$ symmetry with $v_S=0$, the minimum conditions are different to before. This change in minimum conditions modifies many of the expressions previously found. Firstly, the pure $\phi$ couplings would be the same as those in the SM, $m_h^2=2\mu_0^2$ and $\lambda_0=m_h^2/(4v^2)$, since the singlet VEV is zero in the broken phase. This is equivalent to setting $\epsilon=1$ in eq~(\ref{defeps}). Secondly, we can express $b_4$ in terms of the VEV of the singlet field in the symmetric vacuum, $b_4=-b_2/\tilde{v}_S^2$. The vacuum energy difference is given by
\begin{equation}
\Delta V_{\text{1 loop (0T)}}^{\text{[$\mathbb{Z}_2$xSM]}}=-\dfrac{1}{4}m_h^2v^2-\dfrac{1}{4}b_2\tilde{v}_S^2-\dfrac{2}{(16\pi)^2}\left( g_tm_t^4-g_Wm_W^4-g_Zm_Z^4 \right)\text{.}
\label{Z2xSm_raw}
\end{equation}
\noindent Compared to the SM vacuum energy difference there is an extra tree level piece in eq.~(\ref{Z2xSm_raw}), which has the opposite sign to the SM piece if $b_2<0$. In other words, the tree level contribution to the vacuum energy difference will be reduced compared to the SM if $\tilde{v}_S\neq 0$. Since the overall size of this extra term determines the vacuum energy difference, we should investigate this term more closely. Rewriting $b_2$ in terms of the singlet mass and coupling $a_2$,
\begin{equation}
b_2=2m_s^2-a_2v^2\text{,}
\end{equation}
\noindent we find an upper bound for the singlet mass of $m_s^2<a_2v^2/2$. This bound is necessary to decrease $\vert \Delta V_{\text{1 loop (0T)}}^{\text{[$\mathbb{Z}_2$xSM]}} \vert$ compared to the SM value. This implies that in order to have $\tilde{v}_S\neq 0$ and the singlet heavier that the SM Higgs, $m_s > m_h$, we require a relatively large coupling $a_2 \gtrsim1$. From unitarity arguments the maximum value of $a_2$ is about $8\pi$, which translates to an upper bound for the singlet mass of $m_s \sim 600~\text{GeV}$. The singlet mass in the unbroken $\mathbb{Z}_2$ case is given by
\begin{equation}
m_s=\sqrt{\dfrac{1}{2}a_2v^2+\dfrac{2}{\tilde{v}_S^2}\left( \Delta V_{\text{tree}}^{[\text{SM]}}-\Delta V_{\text{tree}}^{[\text{$\mathbb{Z}_2$xSM]}} \right)}\text{.}
\label{msZ2vS0}
\end{equation}
\noindent In contrast to eq.~(\ref{mSZ2}), the singlet mass does not vanish as we take $\Delta V_{\text{tree}}^{[\text{$\mathbb{Z}_2$xSM]}}\rightarrow 0$. In order for the singlet mass to be positive within this limit, it is required that $a_2\tilde{v}_S^2>m_h^2$. Given the maximum value of $a_2\sim 8\pi$, we find that $\vert \tilde{v}_S\vert \gtrsim 25~\text{GeV}$. The $a_2v^2/2$ term in eq.~(\ref{msZ2vS0}) protects the mass of the singlet from vanishing as $\Delta V_{\text{tree}}^{[\text{$\mathbb{Z}_2$xSM]}}\rightarrow 0$. Consequently, the behaviour in taking the vacuum energy difference to zero in the unbroken $\mathbb{Z}_2$ case differs drastically compared to the behaviour in the $\mathbb{Z}_2$ broken case.

\subsection{\label{section:GNMSSMana}The vacuum energy difference (GNMSSM)}
To the tree level potential, we apply the usual minimal conditions to eliminate the $m_{H_u}^2$, $m_{H_d}^2$, and $m_{S}^2$ soft mass parameters in favour of $\tan\beta$ and the VEVs, $v$ and $v_S$. The rest of the analytic work that we concern ourselves with regards the potential in the real singlet direction, $s=\rm{Re}(S)$, defined as the potential at $H_u=H_d=0$. The resulting potential takes the form
\begin{equation}
\begin{array}{l l}
V_{\text{tree (singlet)}}^{\text{[GNMSSM]}}= & \displaystyle k_1^2+[m_S^2+k_2(A_{k_2}+k_2)+k_1k_3]s^2+\dfrac{2}{3}k_3(A_{k_3}+3k_2)s^3+k_3^2s^4\text{,}
\end{array}
\label{GNMSSM_tree_singlet_dir}
\end{equation}
\noindent where we have chosen $A_{k_1}=-k_2$ in order to remove the linear term in this potential without loss of generality\footnote{We can recover an arbitrary value of the chosen parameter by a shift in field $s$.}. Solving for the extremum in the singlet direction, we find a trivial extremum at $s=0$ whose extremum nature depends on the sign of the quadratic term in eq.~(\ref{GNMSSM_tree_singlet_dir}). Note that for a potential bounded from below, we can only have three shapes for the potential in the singlet direction:
\begin{itemize}
  \item Minimum at $s=0$: this is the only extremum.
  \item Minimum at $s=0$: there exist two additional extrema, one maximum and one minimum. The additional minimum having the greater magnitude of $s$.
  \item Maximum at $s=0$: there exist two additional extrema, both minima, whose $s$-values have opposite sign.
\end{itemize}
\noindent In the GNMSSM, we find that the additional extrema are located at
\begin{equation}
\langle s \rangle_\pm=-\dfrac{A_{k_3}+3k_2}{4k_3} \pm \sqrt{\left[ \dfrac{A_{k_3}+3k_2}{4k_3} \right]^2-\dfrac{1}{2k_3^2}\left[ m_S^2+k_2(A_{k_2}+k_2)+2k_1k_3 \right]}\text{.}
\label{S_min_GNMSSM}
\end{equation}
\noindent For three extrema in the singlet direction, this requires the condition that
\begin{equation}
(A_{k_3}+3k_2)^2-8\left[ m_S^2+k_2(A_{k_2}+k_2)+2k_1k_3 \right] \geq 0\text{.}
\end{equation}
\noindent In meeting this condition, assuming small values of $k_3$, and $A_{k_3} \sim k_2 \sim m_{\text{SUSY}}$, there is a strong tendency for an additional minimum to exist at very large singlet field values. This is of course without a tuning of $A_{k_3}$ and $k_2$. It is interesting to note that a tuning to make the ratio $(A_{k_3}+3k_2)/(k_3)$ smaller is analogous to forcing the effective $b_3$ trilinear singlet term (as appears in the xSM model) to be zero. To clarify, we can express the ratio in terms of an effective $b_3$ parameter in place of $A_{k_3}$ and $k_2$
\begin{equation}
\left\vert \dfrac{A_{k_3}+3k_2}{4k_3} \right\vert \sim \left\vert \dfrac{b_3}{8k_3^2} \right\vert.
\label{effCubSUSY}
\end{equation}
\noindent The essential point here is that by capping the additional minimum to less than 10~\text{TeV}, small values of $k_3 < 10^{-3}$ set $\vert b_3 \vert \lesssim 0.1~$\text{GeV}. In contrast, large values of $k_3 \sim 1$ allow for a far larger cubic term, $\vert b_3 \vert \sim 80~$\text{TeV}, but at the risk of other complications to the model. Namely that both $\lambda$ and $k_3$ are large, in tension with theoretical constraints due to the presence of a Landau pole \cite{Ellwanger:1996}. In the numerical analysis, we consider points for the GNMSSM with a cap of 10~\text{TeV} on the field value of all singlet extrema and are thus biased toward a large $\lambda$ and large $k_3$ parameter space.

\section{\label{section:num_scan}Numerical Scan}
By means of a numerical scan over a selected parameter space, we look at various distributions related to the variables $v_c$, $T_c$, $\xi$, and $\Delta V_{\text{1 loop (0T)}}$. The scans are conducted with the aim of covering the range of possibilities. Hence the density of parameter points in the plots is not necessarily representative of a statistical likelihood of landing in any particular region. In our numerical analysis, we vary most of the dimensionful parameters between 0~\text{GeV} and 1000~\text{GeV} to the appropriate power. For details see Appendices~\ref{section:paraxSM}~and~\ref{section:paraGNMSSM}.

\subsection{Phenomenological constraints}
For the $\mathbb{Z}_2$xSM and xSM models, we apply the constraints from \cite{Robens:2015}. This constrains the value of the mixing angle, $\vert\sin\alpha\vert$, against the mass of the singlet, $m_s$. For singlet masses below $80~\text{GeV}$ there is a bound of $\vert\cos\alpha\vert \geq 0.985$ ($\vert\sin\alpha\vert \leq 0.173$). This bound comes from collider exclusion limits, including LHC Higgs signal rates. For singlet masses between $80-180~\text{GeV}$ the mixing angle is constrained by LEP and LHC exclusion bounds. For singlet masses greater than $180~\text{GeV}$, we apply the constraint of quantum corrections to the $W^\pm$ boson mass \cite{Lopez-Val:2014}. We expect the validity of the high singlet mass constraint to breakdown in supersymmetric models due to additional particle content contributing to loop corrections. For the GNMSSM, we instead apply a bound of $\vert\sin\alpha\vert \leq 0.55$ for parameter points with a singlet mass greater than $180~\text{GeV}$ \cite{Jeong:2012}. We cut out stop masses below $m_{\tilde{t}_2} \leq 95.7~\text{GeV}$, in accordance with \cite{Agashe:2014}, but our analysis is not sensitive to this choice.

\subsection{Scan procedure}
We produce random parameter configurations by using flat distributions of the parameters, unless stated otherwise (see Appendices~\ref{section:paraxSM}~and~\ref{section:paraGNMSSM}). We then test if these points pass theoretical and/or phenomenological constraints. These tests are based upon desired features of the one loop zero temperature potential and mass spectrum. All parameter points are subject to theoretically motivated cuts, such as (i) the broken vacuum is the absolute minimum of the one loop zero temperature effective potential, (ii) positivity and non-degeneracy of all physical squared masses, (iii) positivity of the quartic couplings\footnote{In the xSM, this means $\lambda_0,>0$ and $b_4>0$, but $a_2$ can have either sign.}, and (iv) the imaginary singlet direction does not require a VEV.

\subsubsection*{Procedure in the single field model scans}
Starting from the one loop zero temperature potential, we scan over regular intervals of the vacuum energy difference, $\Delta V_{\text{1 loop (0T)}}$, whilst recording the corresponding free parameter of the model. Initially taking the minimum and maximum temperature to be 0~\text{GeV} to 200~\text{GeV} respectively, we use a simple algorithm to iteratively change the minimum/maximum temperature. The temperatures are updated according to whether the broken vacuum is higher or lower than the symmetric vacuum at the temperature midway between the minimum/maximum temperatures in the current iteration. The final VEV of $\phi$ and temperature are recorded as the critical values for each parameter point.

\subsubsection*{Procedure in the xSM}
Since the algebraic form of the one loop zero temperature vacuum energy difference is generally quite complicated, we adopt a semi-analytic approach to study this model. Rather than scanning over regular intervals of the one loop vacuum energy difference, we perform a random scan over the free parameters and rely on a numerical analysis to ensure the potential is theoretically well-behaved, i.e. bounded from below with the broken vacuum as the absolute minimum. Our numerical work confirms that the expressions for the vacuum energy difference in Section~\ref{section:xSMvac} are correct.

For the $\mathbb{Z}_2$ case, we also randomly assign values to the free parameters in accordance with the ranges in Table~\ref{xSM_para_choice} found in Appendix~\ref{section:paraxSM}. For the unbroken $\mathbb{Z}_2$ case, $\lambda_0$ is fixed by $m_h$, and rather than reparameterising, we scan over the remaining quartics, $a_2$ and $b_4$, as well as the singlet mass, $m_s$.

\subsubsection*{Procedure in the GNMSSM}
This model is investigated through an almost entirely numerical manner. The parameter scan sequentially performs checks at tree, one loop zero temperature, and one loop finite temperature level.
\begin{enumerate}
  \item \textbf{Tree level parameter point scan:}
  \begin{enumerate}[(a)]
    \item Randomly assign a numerical value to the tree level parameters, in accordance with Table~\ref{rand_para_choice} in Appendix~\ref{section:paraGNMSSM}.
    \item Find $A_{k_1}$ and $A_{\lambda}$ such that (i) no linear singlet term exists in the potential (we find $A_{k_1}=-k_2$ is always the case at tree level) and (ii) that the broken vacuum is lower than the minimum value in the singlet direction.
    \item Check the mass spectrum of the Higgs sector. Pass any points that find (i) the $h^0$ state with mass between $0.5 \times 125~\text{GeV}$ and $125~\text{GeV}$, (ii) the $H^0$, $A^0$, and $H^\pm$ states have masses exceeding $200~\text{GeV}$, and (iii) both singlet-dominant states are positive in mass.
  \end{enumerate}

  \item \textbf{One loop zero temperature parameter point scan:}
  \begin{enumerate}[(a)]
    \item To reduce the number of parameters, we choose the off-diagonal terms of the stop squared-mass matrix to be zero at the minimum. This means $A_t=(\mu+\lambda v_S)\cot\beta$. Furthermore, for the diagonal terms we take $m_{Q_3}=m_{\bar{u}_3}+\Delta m_3$, where $\Delta m_3=100~\text{GeV}$. The phase transition is not effected by these choices since it is not induced by light stops.
    \item Given the range $\Delta m_3 < m_{Q_3} \leq m_{\text{SUSY}}$, we perform a simple scan over $m_{Q_3}$. until the stop contribution to the one loop potential results in a numerical value of $m_{h^0}=125~\text{GeV}$. All points find $m_{h^0}$ accurate to within $\pm 0.02~\text{GeV}$.
  \end{enumerate}
  
  \item \textbf{One loop finite temperature parameter point scan:}
  \begin{enumerate}[(a)]
    \item Numerically scan over the temperature between 0~\text{GeV} and 200~\text{GeV}, finding the broken and symmetric vacua at each temperature.
    \item Reiterate the above step multiple times, closing in on the temperature at which the vacuum energy difference is zero. Record the critical temperature, $T_c$, and critical field values, $v_c$, $(\tan\beta)_c$, $(v_S)_c$, and $(\tilde{v}_S)_c$.
  \end{enumerate}
\end{enumerate}

\subsection{\label{section:scatPlots}Numerical results}
Let us discuss the main qualitative features of the numerical results. These features are best captured by Figures~\ref{singleField_PtPlot_fig}, \ref{Z2xSM_PtPlot_fig}, \ref{xSM_PtPlot_fig}, and \ref{GNMSSM_PtPlot_fig}. All of these figures show that an increase in the vacuum energy difference at one loop zero temperature increases the strength of the phase transition. However, the precise relation between the strength of the phase transition and the vacuum energy difference requires a detailed investigation.

\subsubsection{\label{section:numSingle}Single field models}
\begin{figure}[t]
\centering
\includegraphics[width=15.0cm]{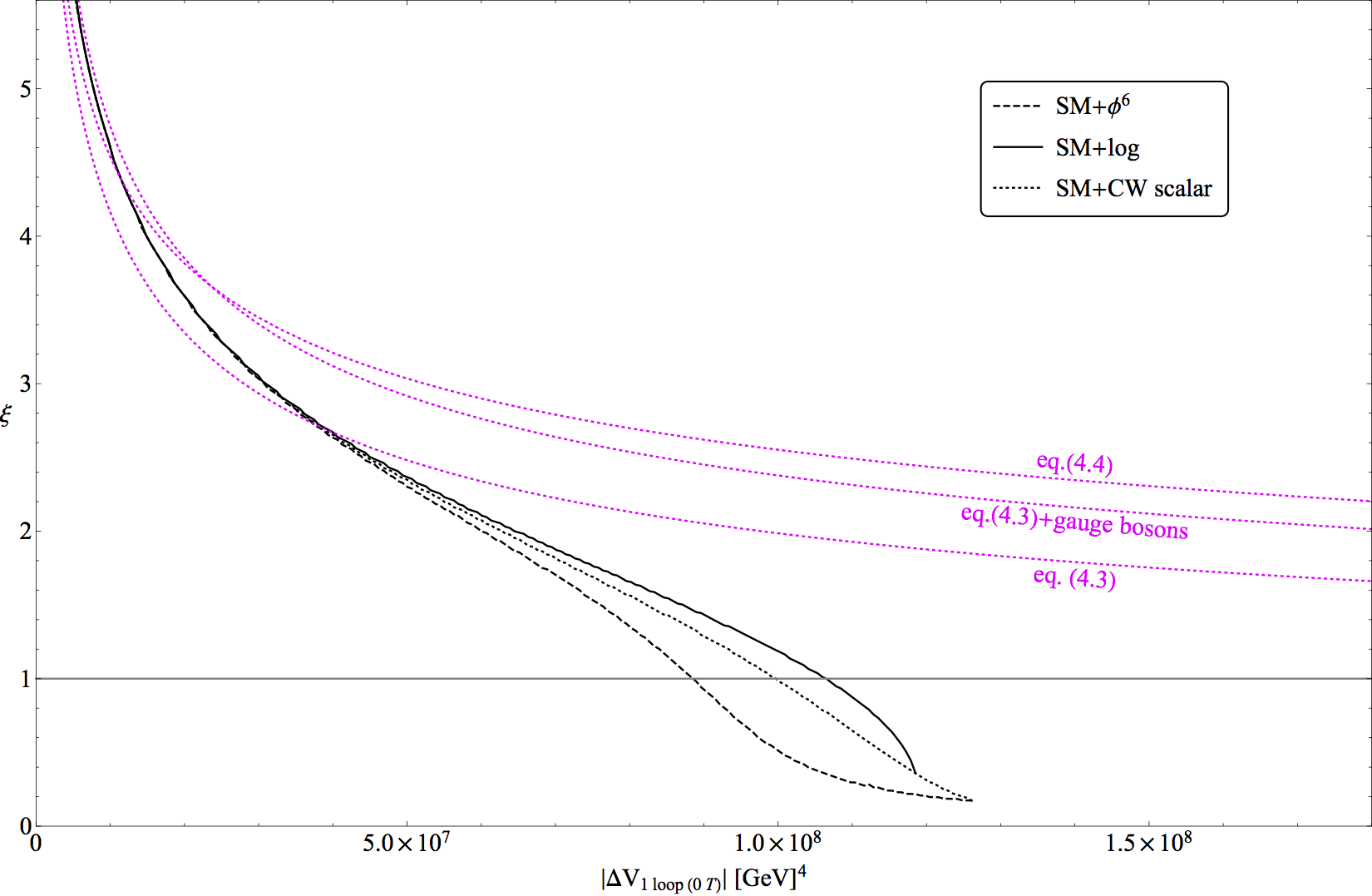}
\caption{\small{Strength of the phase transition, $\xi$, against the magnitude of the vacuum energy difference, $\vert\Delta V_{\text{1 loop (0T)}}\vert$, for the single field models. The magenta curves display the prediction for the strength $\xi$ when the broken minimum is considered in a low temperature expansion. Also shown is the $\xi$=1 line.}}
\label{singleField_PtPlot_fig}
\end{figure}

For the single field models investigated, we can understand that the strength of the phase transition $\xi$ increases as a result of two effects. The first is that the broken vacuum at critical temperature remains close to its zero temperature field VEV. The second is that the critical temperature decreases with the magnitude of the vacuum energy difference. So in the limit $\vert \Delta V_{\text{1 loop (0T)}}\vert \rightarrow 0$,
\begin{equation}
v_c\rightarrow v\text{ and }T_c\rightarrow 0 \Rightarrow \xi \rightarrow \infty\text{.}
\end{equation}
Clearly one would expect metastability of the symmetric phase in the limit of large $\xi$, but this is not the focus of the current discussion. One interesting observation from Figure~\ref{singleField_PtPlot_fig} is that there exists a universal behaviour at low values of $\vert \Delta V_{\text{1 loop (0T)}}\vert$. To understand the reason for such behaviour we need an expression for the strength at low critical temperature values.

In order to determine an analytic form for the strength of the phase transition we must take care to use the correct analytic limit for the thermal potential. In the cases we investigate, the high temperature expansion is always valid in the symmetric vacuum. In terms of the dynamics of increasing temperature, the value of the potential in the symmetric vacuum is shifted proportional to $T^4$. However, in a neighbourhood of the broken vacuum, we are in a low temperature regime. In the low temperature limit, the thermal contribution to the potential is given by \cite{Dolan:1973}
\begin{equation}
V_{\text{low T}}(\phi,T)= \sum_{i=f,b}^\infty g_iT^4\left( \dfrac{m_i(\phi)}{2\pi T} \right)^{3/2}\exp\left( -\dfrac{m_i(\phi)}{T} \right)\left( 1+\dfrac{15}{8}\dfrac{T}{m_i(\phi)} \right)\text{.}
\label{lowTV}
\end{equation}
In the cases we consider in Figure~\ref{singleField_PtPlot_fig}, the top quark contribution dominates the expression in eq.~(\ref{lowTV}) and so we will neglect the contribution from the EW gauge bosons. Since the vacuum energy difference is zero at the critical temperature, one may equate the required thermal contribution to the vacuum energy difference with the zero temperature value. Assuming $v_c\approx v$ for parameter regions with a low critical temperature, we can derive an equation for $\xi$ as follows
\begin{equation}
\dfrac{\vert \Delta V_{\text{1 loop (0T)}} \vert}{4v^4}\approx g_t \xi^{-4}\left[ \dfrac{7\pi^2}{720}-\left( \dfrac{y_t}{\sqrt{2}}\dfrac{\xi}{2\pi} \right)^{3/2}\exp\left( -\dfrac{y_t}{\sqrt{2}}\xi \right)\left( 1+\dfrac{15}{8}\dfrac{\sqrt{2}}{y_t \xi} \right)\right]\text{.}
\label{lowTsat}
\end{equation}
Taking the limit that the strength $\xi$ is very large, the exponential term suppresses all $\xi$-dependent terms inside the square bracket in eq.~(\ref{lowTsat}). Then the strength of the phase transition is estimated to be
\begin{equation}
\xi\approx \sqrt{2}v\left( \dfrac{7\pi^2}{720}\dfrac{g_t}{\vert \Delta V_{\text{1 loop (0T)}} \vert} \right)^{1/4}\text{.}
\label{lowTstrong}
\end{equation}

These approximations are shown as dotted lines in Figure~\ref{singleField_PtPlot_fig} and reproduce the full result reasonably well for large values of $\xi$. As $\xi$ becomes larger than about 5, also the gauge bosons will reach a low temperature regime in the broken phase and should be included. Adding them in eq.~(\ref{lowTsat}) leads to a very accurate estimate labelled as ``eq.~(\ref{lowTsat})+gauge bosons'' in Figure~\ref{singleField_PtPlot_fig}. So the observed universal behaviour is fixed by the number of relevant degrees of freedom in the plasma. These are the particles which become massless in the symmetric phase and Boltzmann suppressed in the broken phase. Finally, we can use eq.~(\ref{lowTstrong}) to derive a simple estimate for the critical temperature,
\begin{equation}
T_c\approx\left( \dfrac{720}{7\pi^2 g_t}\vert \Delta V_{\text{1 loop (0T)}}\vert\right)^{1/4}.
\label{lowTc}
\end{equation}

\begin{table}[b]
\centering
\begin{tabular}{c | c c c c c c}
\hline
\textbf{Model:} & & SM$+\phi^6$ & SM+log & SM+CW \text{scalar}\\
\hline
\textbf{Free parameter:} & & $M$ & $\lambda_0$ & $y$ \\
\textbf{Bound:} & & $<854~\text{GeV}$ & $>0.142$ & $>2.47$ \\
\hline
\end{tabular}
\caption{\small{Bounds on the free parameters in the single field models that guarantee a strong phase transition.}}
\label{single_bounds}
\end{table}

\noindent In order to guarantee a strong phase transition for each of the single field modifications to the SM, we find bounds on each of the free parameters (see Table~\ref{single_bounds}). For the SM with a dimension-six operator, the mass suppression favouring a low scale cutoff has been studied in ref.~\cite{Grojean:2004,Bodeker:2004}. These translate as upper bounds on the vacuum energy difference of
\begin{equation}
\vert \Delta V_{\text{1 loop (0T)}} \vert < \left\{
\begin{array}{ l l }
8.83 \times 10^7~\text{GeV}^4 & \text{for the SM}+\phi^6\text{,}\\
1.06 \times 10^8~\text{GeV}^4 & \text{for the SM}+\text{log}\text{,}\\
9.95 \times 10^7~\text{GeV}^4 & \text{for the SM}+\text{CW scalar}\text{.}
\end{array}\right.
\label{singleBound}
\end{equation}
Each hints at the necessity for below TeV scale physics and additional scalar states/extended Higgs sectors. It is interesting to note that a very mild modification of the vacuum energy by about 25\% is sufficient to induce a strong first order transition. 

\subsubsection{\label{section:numNonSUSY}Non-supersymmetric singlet extension}
\begin{figure}[t]
\centering
\begin{minipage}[b]{7.5cm}
\centering
\subcaptionbox{\small{Without phenomenological constraints.}}{\includegraphics[width=7.5cm]{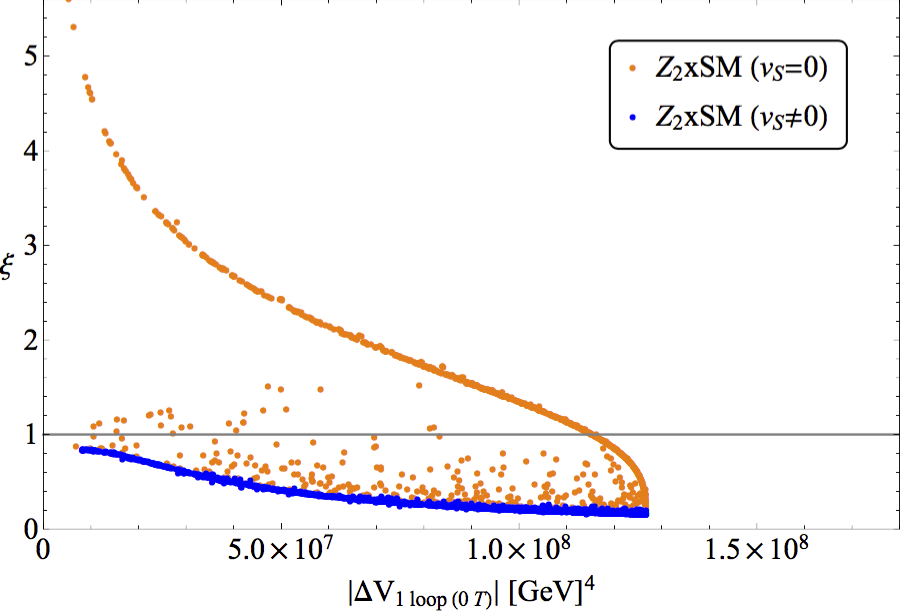}}
\end{minipage}
\begin{minipage}[b]{7.5cm}
\centering
\subcaptionbox{\small{With phenomenological constraints.}}{\includegraphics[width=7.5cm]{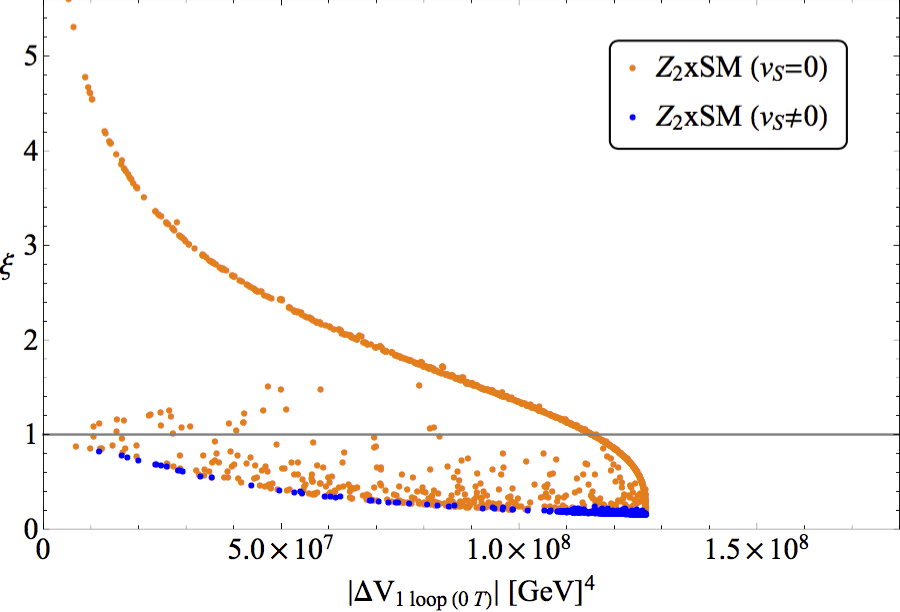}}
\end{minipage}
\caption{\small{Strength of the phase transition, $\xi$, against the magnitude of the vacuum energy difference, $\vert\Delta V_{\text{1 loop (0T)}}\vert$, for the $\mathbb{Z}_2$xSM ($v_S=0$) and $\mathbb{Z}_2$xSM ($v_S\neq0$) singlet extensions. Also shown is the $\xi$=1 line.}}
\label{Z2xSM_PtPlot_fig}
\end{figure}
Next we will remark on Figure~\ref{Z2xSM_PtPlot_fig}, which shows parameter points for the $\mathbb{Z}_2$xSM, where the $\mathbb{Z}_2$ symmetry is either spontaneously broken or unbroken at zero temperature. The universal behaviour seen in Figure~\ref{singleField_PtPlot_fig} is also observed for a number of parameter points in the unbroken case. However, there are some parameter points that do not follow this universal curve and instead fall somewhere between this curve and another branch. This other branch happens to be traced out by all points in the spontaneously broken case. Unfortunately, this second branch fails to meet the hypothesis that the phase transition becomes strong (let alone arbitrarily strong) as $\vert \Delta V_{\text{1 loop (0T)}}\vert$ is decreased.

This second branch exists because the second derivative of the broken vacuum changes sign in one direction as the potential is thermally evolved to the critical temperature. This is to say that we lose control over the broken vacuum and it no longer remains close to its zero temperature location in field space. Instead the broken vacuum slides quickly across field space upon small changes in temperature. In such scenarios, we observe that the broken vacuum always slides toward the symmetric phase as the temperature is increased. This sliding of the broken vacuum is analogous to saying that the barrier between the symmetric and broken vacua virtually disappears. The only barrier remaining is that generated through the cubic terms of the EW gauge bosons. The phase transition is therefore SM-like with the physical Higgs mass replaced by its value at $\phi=0$ and $S=\tilde{v}_S$. To avoid such scenarios, one must ensure that the Higgs squared mass matrix is always positive in a neighbourhood of the broken vacuum. The size of this neighbourhood has to be larger if the critical temperature is higher, because then the broken minimum moves more in field space under thermal effects. Therefore, we revise our original statement in Section~\ref{section:vac}:
\begin{center}
\textit{The smaller the value of }$\vert \Delta V_{\text{1 loop (0T)}}\vert$\textit{, the lower the critical temperature. Further, the strength of the phase transition $\xi$ will become arbitrarily strong so long as the Higgs squared mass matrix remains positive in the neighbourhood of the broken vacuum.}
\end{center}

{
Let us stress again that in the current work we choose to use the one loop approximation to the effective potential. In some models the tree level approximation will be sufficient to indicate a first order phase transition, while in other models higher loop orders will have non-negligible impact and need to be included.}

\begin{figure}[t]
\centering
\begin{minipage}[b]{7.5cm}
\centering
\subcaptionbox{\small{Fixed coupling, $a_2=1.0$.}}{\includegraphics[width=7.5cm]{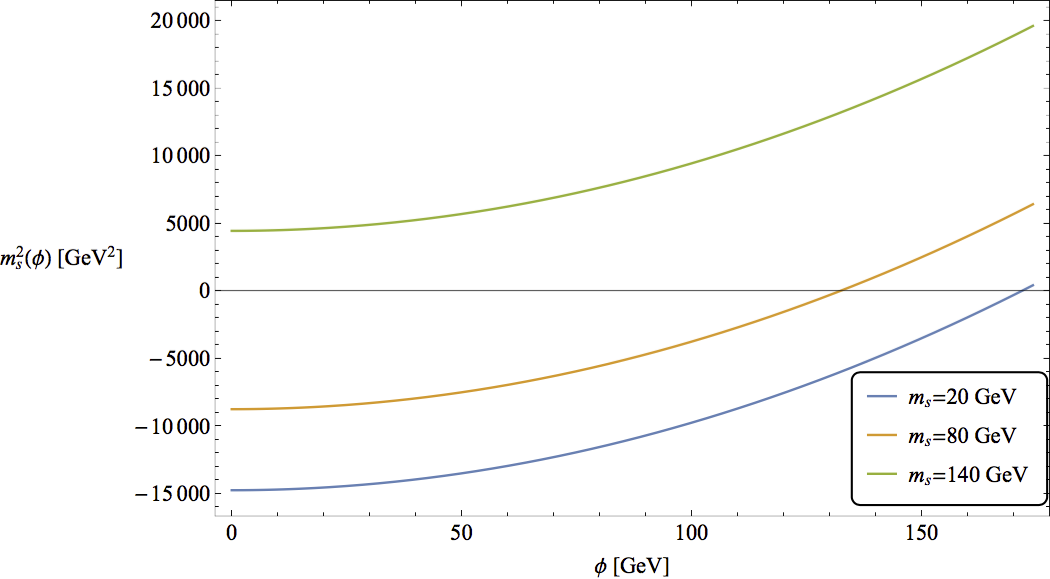}}
\end{minipage}
\begin{minipage}[b]{7.5cm}
\centering
\subcaptionbox{\small{Fixed singlet mass, $m_s=80$~GeV.}}{\includegraphics[width=7.5cm]{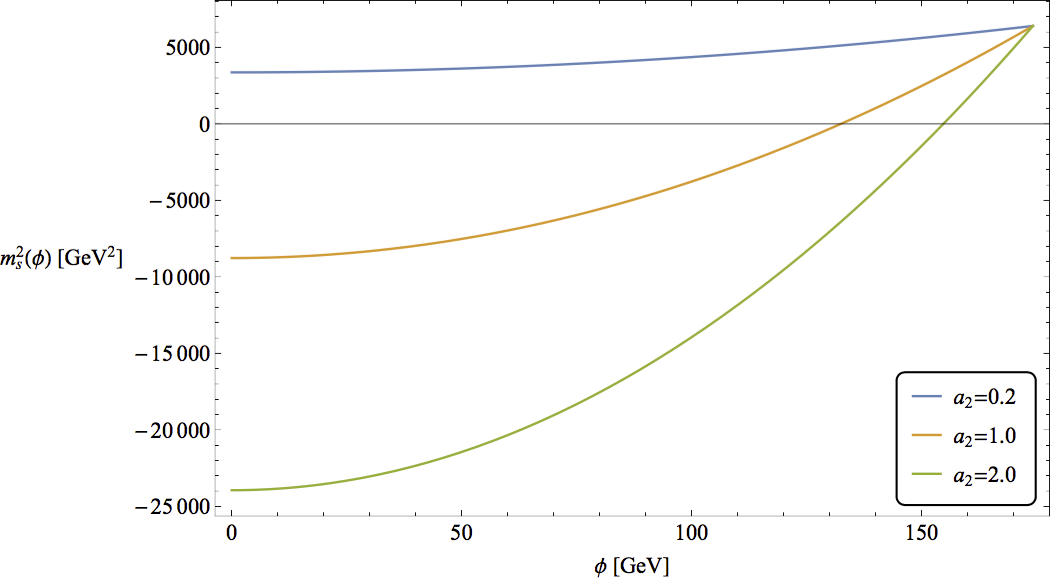}}
\end{minipage}
\caption{\small{Plot of the field-dependent singlet mass at $S=0$ against the $\phi$ direction for various values of $m_s$ and $a_2$ in the $\mathbb{Z}_2$xSM (unbroken). The $\mathbb{Z}_2$ symmetry spontaneously breaks at the value of $\phi$ where the singlet mass squared changes sign. The value of $m_s$ controls the offset of the singlet mass away from $m_s(\phi)=0$. For a given value of $a_2$, a lighter singlet mass brings the $\mathbb{Z}_2$ breaking critical field value closer to the zero temperature VEV, $v$. For a given value of $m_s$, the higher the value of the quartic coupling $a_2$, the closer the $\mathbb{Z}_2$ breaking critical field value is to the zero temperature VEV, $v$.}}
\label{Z2xSM_vary}
\end{figure}

Let us consider the case where the $\mathbb{Z}_2$ symmetry is unbroken at zero temperature. Parameter points that undergo spontaneous $\mathbb{Z}_2$ breaking between zero temperature and the critical temperature are those observed either between the two branches in Figure~\ref{Z2xSM_PtPlot_fig} or lie on the same branch as the parameter points in the $\mathbb{Z}_2$ broken at zero temperature case. The points on the ``universal'' branch remain unbroken up to the critical temperature.

For the case where the $\mathbb{Z}_2$ symmetry is unbroken at zero temperature, the field-dependent singlet mass at $S=0$ is given by
\begin{equation}
m_s^2(\phi)=m_s^2+\dfrac{a_2}{2}(\phi^2-v^2)\text{,}
\label{singletfielddep}
\end{equation}
where $m_s$ is the mass of the singlet at $\phi=v$. Figure~\ref{Z2xSM_vary} shows how the critical Higgs field value (where the $\mathbb{Z}_2$ symmetry breaks) depends on the zero temperature quantities $m_s$ and $a_2$. To avoid the $\mathbb{Z}_2$ symmetry breaking due to thermal effects, we must ensure that the mass-squared value of the singlet remains positive in the broken minimum up to the critical temperature. One may thus always guarantee a strong phase transition using our hypothesis by choosing $m_s$ and $a_2$ such that eq.~(\ref{singletfielddep}) is positive. A sliding singlet occurs for a light singlet mass and large $a_2$ coupling. In these cases, the small singlet mass results from a more or less severe tuning between bare and electroweak symmetry breaking induced terms.

It should be noted that phenomenological constraints only apply at zero temperature. Therefore all parameter points in the $\mathbb{Z}_2$ unbroken case are viable candidates for a theory beyond the SM, since there is no Higgs-singlet mixing at zero temperature. However, a spontaneous breaking of the $\mathbb{Z}_2$ symmetry before the start of the electroweak phase transition disfavours a strong phase transition. A more striking observation is that if the $\mathbb{Z}_2$ is spontaneously broken at zero temperature, then no points achieve a strong phase transition. This may be slightly modified by thermal effects, e.g. an enhancement of the thermally-induced barrier when the Higgses are included. Let us also note that in the case of spontaneous $\mathbb{Z_2}$ breaking, phenomenological constraints remove most of our parameter sets. So spontaneous $\mathbb{Z}_2$ breaking before the critical temperature is phenomenologically disfavoured and, if realised, does not lead to a strong phase transition. This observation is consistent with the findings in ref. \cite{Profumo:2007,Espinosa:2011,Barger:2011}.

\begin{figure}[t]
\centering
\begin{minipage}[b]{7.5cm}
\centering
\subcaptionbox{\small{Without phenomenological constraints.}}{\includegraphics[width=7.5cm]{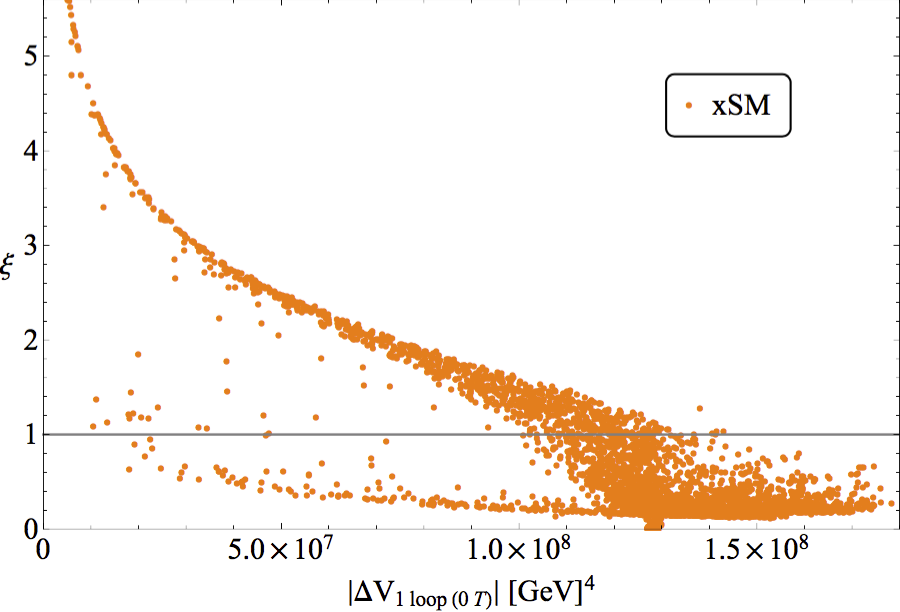}}
\end{minipage}
\begin{minipage}[b]{7.5cm}
\centering
\subcaptionbox{\small{With phenomenological constraints.}}{\includegraphics[width=7.5cm]{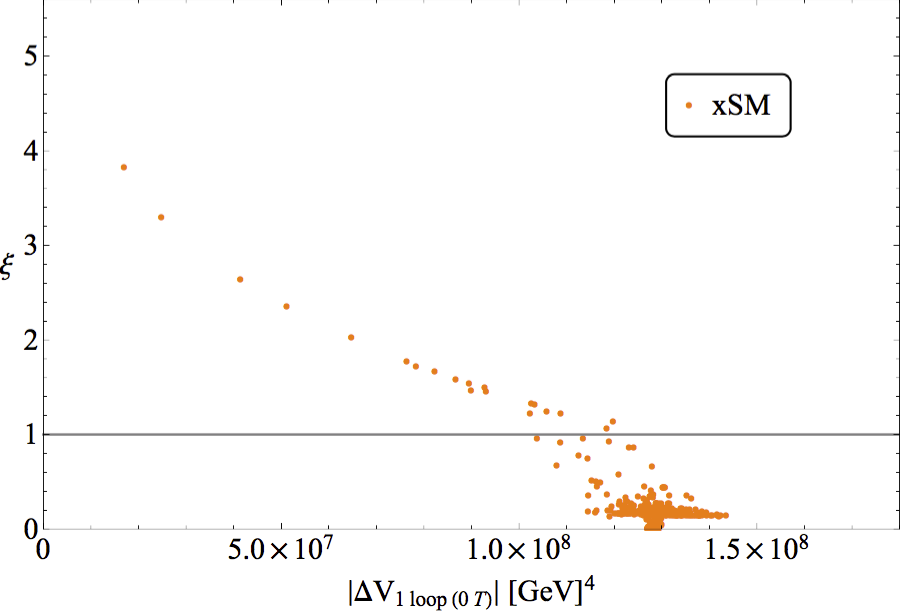}}
\end{minipage}
\caption{\small{Distribution of the strength of the phase transition, $\xi$, against the magnitude of the vacuum energy difference, $\vert\Delta V_{\text{1 loop (0T)}}\vert$, for the xSM with the $\mathbb{Z}_2$ explicitly broken at zero temperature. Also shown is the $\xi$=1 line.}}
\label{xSM_PtPlot_fig}
\end{figure}

Let us now turn to the xSM with the $\mathbb{Z}_2$ explicitly broken at zero temperature. The parameter points for this model can be found in Figure~\ref{xSM_PtPlot_fig}. In comparison with the $\mathbb{Z}_2$xSM cases in Figure~\ref{Z2xSM_PtPlot_fig}, we observe identical behaviour including the universal behaviour at low $\vert \Delta V_{\text{1 loop (0T)}}\vert$. As for the physics, the main qualitative difference between the xSM and $\mathbb{Z}_2$xSM is that the $\mathbb{Z}_2$ is explicitly broken rather than \textit{possibly} spontaneously broken. An interesting contrast between the xSM and $\mathbb{Z}_2$xSM (broken) case is that a lot of parameter points in the xSM do follow our hypothesis. This suggests that for a strong phase transition and a non-zero Higgs-singlet mixing at zero temperature, the potential must contain non-thermal cubic terms for our hypothesis to succeed. In support of this statement, we find that all parameter points on the undesirable branch (traced by $\mathbb{Z}_2$xSM ($v_S\neq 0$) in Figure~\ref{Z2xSM_PtPlot_fig}) vanish if we demand a large cubic term, $a_1>250~\text{GeV}$. We also observe that phenomenological constraints remove the majority of parameter points. Those surviving strictly follow our hypothesis that a tuning of the vacuum energy difference leads to a strong phase transition. After imposing phenomenological constraints, a strong phase transition is guaranteed if $\vert \Delta V_{\text{1 loop (0T)}} \vert< 1.03 \times 10^8~\text{GeV}^4$, i.e.~again a 25\% tuning in the vacuum energy is sufficient.

These results are consistent with the findings of ref.~\cite{Espinosa:2011}. The only exception is that we have not found any parameter points with a strong phase transition in the one loop $\mathbb{Z}_2$xSM ($v_S \neq 0$) model. This very feature was noted in \cite{Espinosa:2011} as being contradictory to other literature, such as \cite{Barger:2007}. We have identified that the $\mathbb{Z}_2$xSM with and without the $\mathbb{Z}_2$ symmetry broken are completely different physical scenarios. This is because the unbroken case does not mix the SM-like Higgs and singlet, whereas the broken case allows for arbitrary mixing. In the unbroken case, a strong phase transition is much more natural to realise.

\subsubsection{GNMSSM}
\begin{figure}[t]
\centering
\begin{minipage}[b]{7.5cm}
\centering
\subcaptionbox{\small{Without phenomenological constraints.}}{\includegraphics[width=7.5cm]{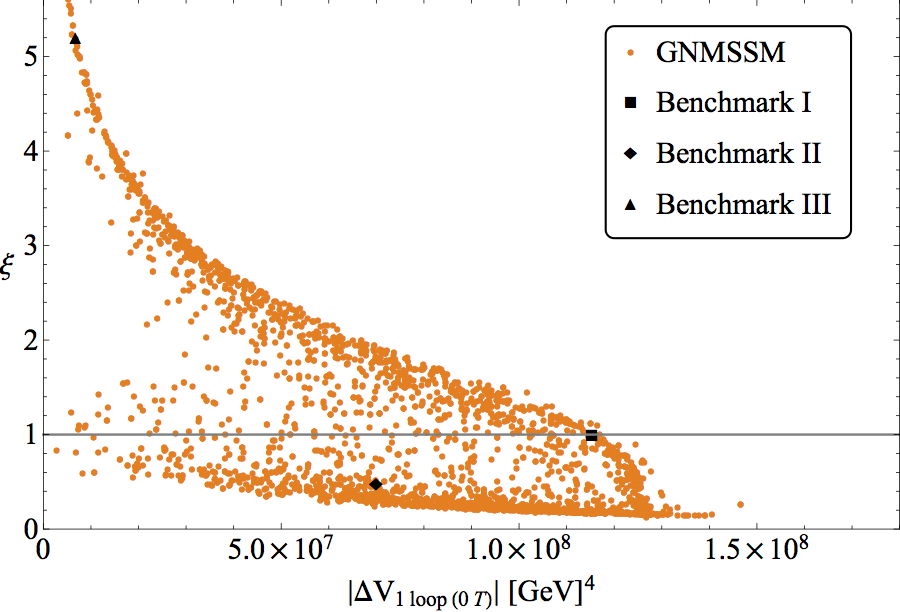}}
\end{minipage}
\begin{minipage}[b]{7.5cm}
\centering
\subcaptionbox{\small{With phenomenological constraints.}}{\includegraphics[width=7.5cm]{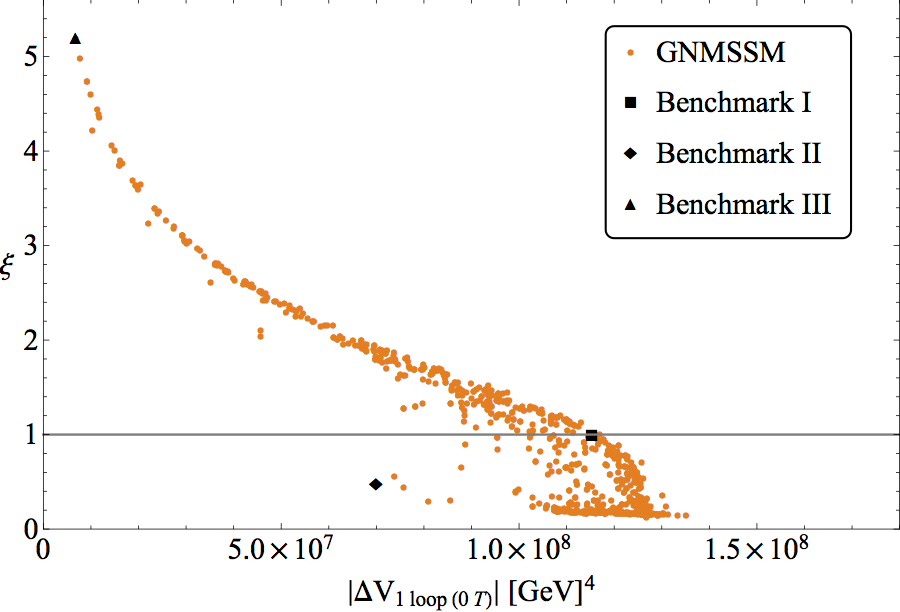}}
\end{minipage}
\caption{\small{Distribution of the strength of the phase transition, $\xi$, against the magnitude of the vacuum energy difference, $\vert\Delta V_{\text{1 loop (0T)}}\vert$, for the GNMSSM. The three benchmark models, chosen from the GNMSSM data set and discussed in Section~\ref{section:bench}, are marked above. Also shown is the $\xi$=1 line.}}
\label{GNMSSM_PtPlot_fig}
\end{figure}

Let us now turn to the GNMSSM. Comparing Figures~\ref{xSM_PtPlot_fig} and \ref{GNMSSM_PtPlot_fig} there is little difference between the GNMSSM and the non-supersymmetric singlet extended cases. However, we notice that the GNMSSM parameter points are more dispersed between the two branches. We suspect that this is because our scanning procedure happens to capture some of the more finely-tuned parameter regions of the supersymmetric theory. This is apparent when we look at the tree level expression for the singlet mass at $S=0$,
\begin{equation}
m_s^2(\phi) = m_s^2+\lambda (\lambda-k_3\sin2\beta)(\phi^2-v^2)\text{,}
\end{equation}
which is the GNMSSM analog of eq.~(\ref{singletfielddep}). Unlike in the xSM where we perform a scan over potentially large values of the $a_2$ coupling through eq.~(\ref{a1xSM}), we are forced in the GNMSSM to keep the $\lambda$ value small to avoid running into a Landau pole \cite{Ellwanger:1996}. These couplings are crucial since they control the second derivative of the singlet field-dependent mass at $S=0$, and hence the chance of finding a parameter point where the potential is destabilised in the singlet direction. An example of such a situation is given as benchmark II discussed below, with related Figure~\ref{benchII}, in Section~\ref{section:bench}. Like in the general xSM, many parameter points are excluded by phenomenological constraints. In particular, because of too large of a Higgs-singlet mixing. For the remaining points, there is a clear relationship between the vacuum energy difference and the strength of the phase transition $\xi$. Our estimates for the strength of the phase transition, eq.~(\ref{lowTstrong}), and critical temperature, eq.~(\ref{lowTc}), still apply.

\begin{figure}[t]
\centering
\includegraphics[width=15.0cm]{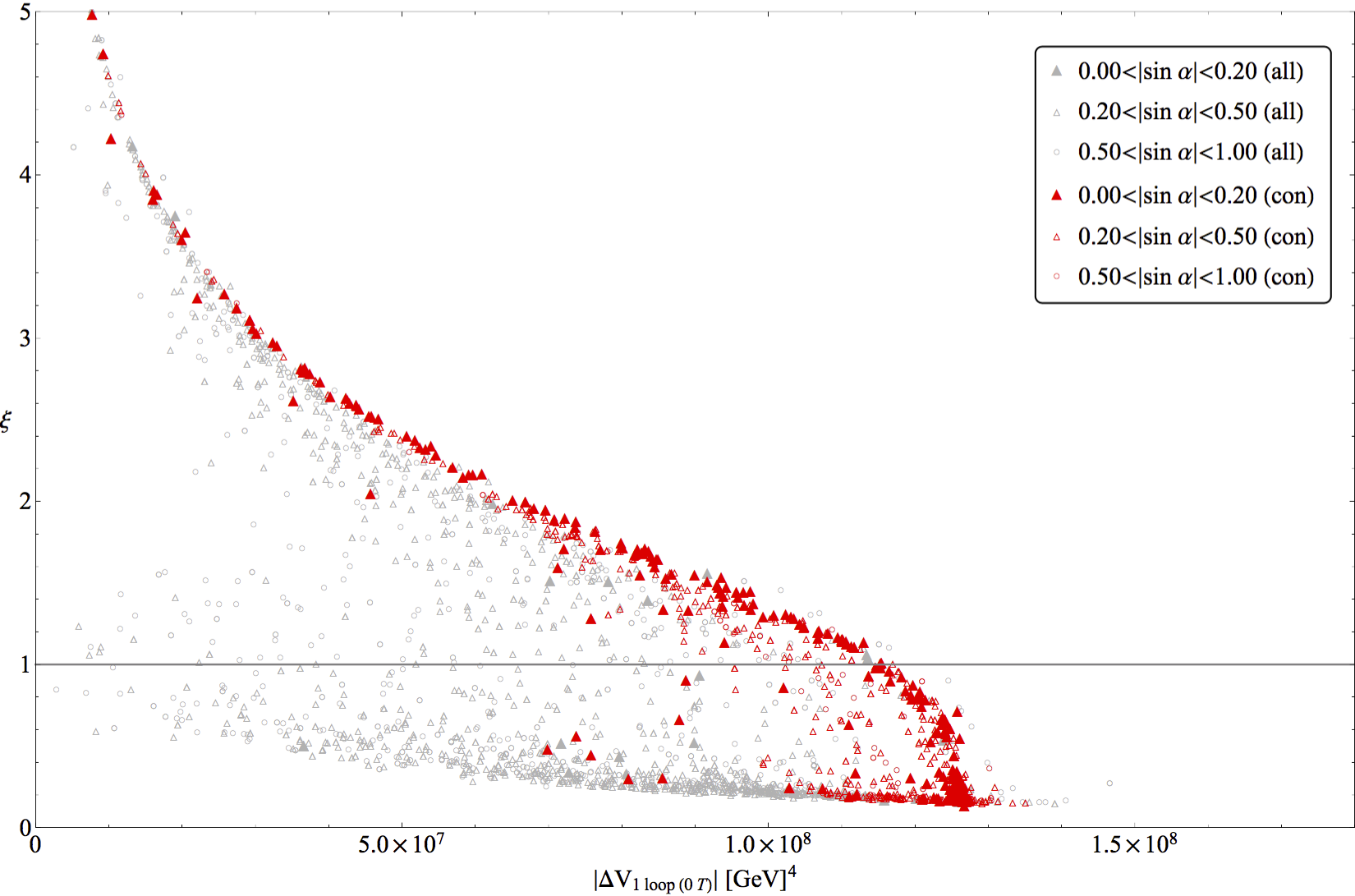}
\caption{\small{Distribution of the strength of the phase transition, $\xi$, against the magnitude of the vacuum energy difference, $\vert\Delta V_{\text{1 loop (0T)}}\vert$, for the GNMSSM with the mixing shown. Note in the key: (all) denotes all of the parameter points and (con) denotes the parameter points that satisfy phenomenology constraints.}}
\label{GNMSSM_PtPlot_mixing_fig}
\end{figure}

Interestingly, we observe a tendency for points with small mixing, $\vert \sin\alpha \vert <0.2$, to lead to a strong $\xi$-$\vert \Delta V_{\text{1 loop (0T)}} \vert$ correlation, as can be seen from Figure~\ref{GNMSSM_PtPlot_mixing_fig}. Similar findings are reported in ref. \cite{Huang:2014} which covers the NMSSM in the limit of no mixing, i.e. $\vert \sin\alpha\vert \rightarrow 0$.

\begin{figure}[t]
\centering
\includegraphics[width=15.0cm]{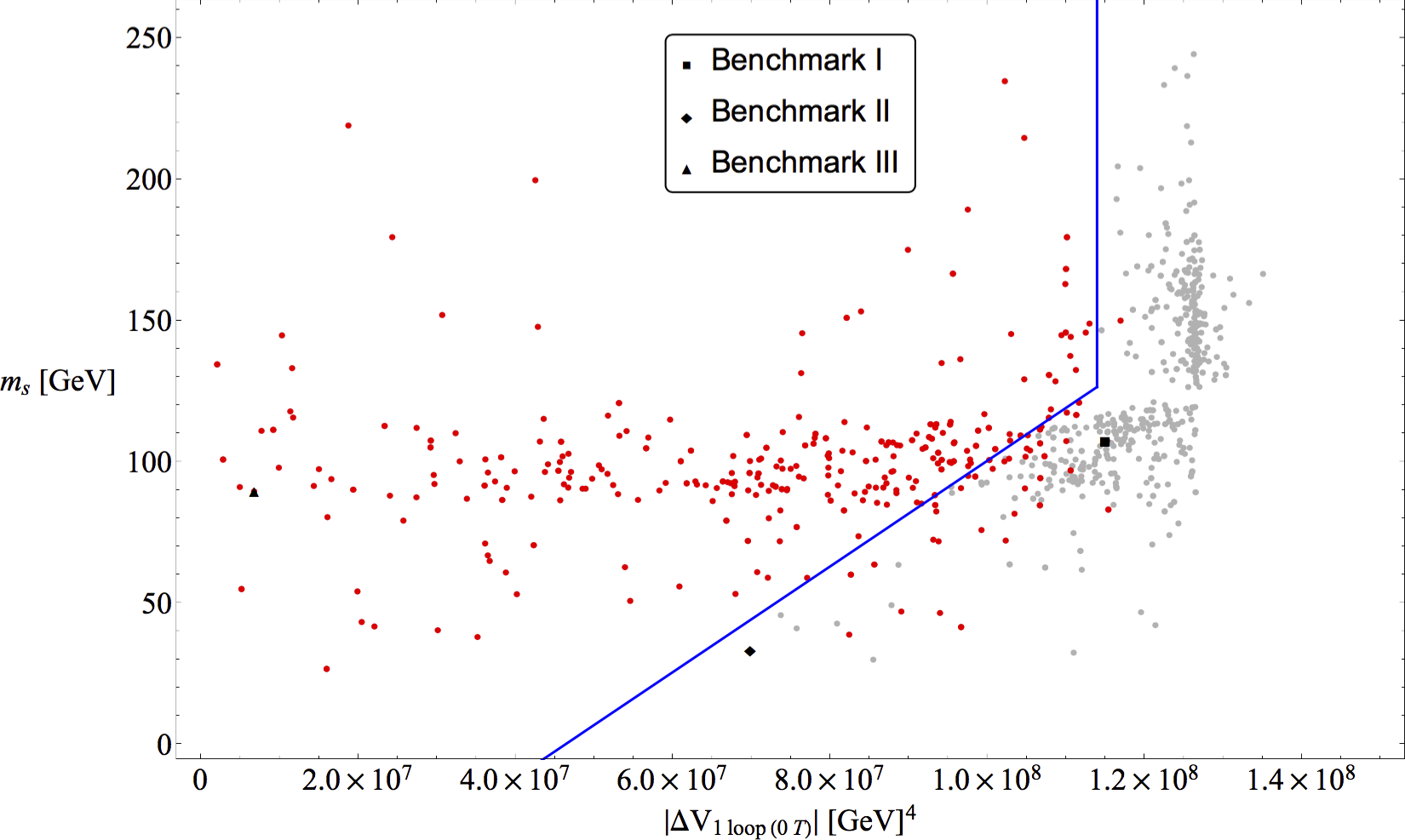}
\caption{\small{Plot of the singlet mass, $m_s$, against the vacuum energy difference, $\vert \Delta V_{\text{1 loop (0T)}}\vert$, for the GNMSSM data set with phenomenological constraints applied. Parameter points highlighted in red have a strong phase transition ($\xi>1$), all other points do not ($\xi<1$). The blue line indicates the bound suggested in eq.~(\ref{recipepheno}).}}
\label{GNMSSMms0SFOpheno_fig}
\end{figure}

For the data set with phenomenological constraints applied we can see an upper bound of $\vert \Delta V_{\text{1 loop (0T)}} \vert< 6.98 \times 10^7~\text{GeV}^4$ ensures we have a strong phase transition. However, this bound removes a significant portion of our parameter space with a strong phase transition. In order to capture more parameter points with a strong phase transition, we instead impose the simultaneous constraints
\begin{equation}
\begin{matrix}
m_s > (87.1~\text{GeV})\times\left(\dfrac{\vert \Delta V_{\text{1 loop (0T)}} \vert}{4.65\times 10^7~\text{GeV}^4}-1\right)\\
\text{and }\vert \Delta V_{\text{1 loop (0T)}}\vert<1.14\times 10^8~\text{GeV}^4\text{.}
\end{matrix}
\label{recipepheno}
\end{equation}

\noindent This bound is indicated in Figure~\ref{GNMSSMms0SFOpheno_fig}, where it is clear that a significant number of points with a strong phase transition are captured. It should be stressed that the recipe in eq.~(\ref{recipepheno}) is only applicable to the GNMSSM with phenomenological constraints applied. Without phenomenological constraints applied a significant number of points with a weak phase transition (many small singlet masses with large Higgs-singlet mixing) appear in the parameter space covered by eq.~(\ref{recipepheno}). For the raw data set, we suggest a modified bound of
\begin{equation}
\begin{matrix}
m_s > (116~\text{GeV})\times \left(\dfrac{\vert \Delta V_{\text{1 loop (0T)}} \vert}{1.14 \times 10^8~\text{GeV}^4}\right)^{1/2}\\
\text{and }\vert \Delta V_{\text{1 loop (0T)}}\vert<1.14\times 10^8~\text{GeV}^4\text{.}
\end{matrix}
\label{recipe}
\end{equation}
A similar bound may be found for the non-supersymmetric models. Note how benchmark III comfortably sits within this territory whereas both benchmarks I and II would be excluded by eq.~(\ref{recipepheno}).

In summary, we find that after applying phenomenological constraints a strong first order phase transition in the GNMSSM requires (modest) tuning of the vacuum energy difference by around roughly 30\%, i.e. from $-1.3 \times 10^8~\text{GeV}^4$ to $-0.9 \times 10^8~\text{GeV}^4$. This is not a significant amount of tuning. So a strong first order phase transition is easily realisable in the context of this model.

\subsection{\label{section:bench}GNMSSM benchmark models}
Here we will look at three benchmarks in our GNMSSM data set that satisfy phenomenological constraints. We have chosen the benchmarks based on the strength of the phase transition $\xi$ and the value of the vacuum energy difference. All three are indicated in Figures~\ref{GNMSSM_PtPlot_fig} and \ref{GNMSSMms0SFOpheno_fig}. More specifically, we choose benchmark I (benchmark III) to have a strong phase transition but large (small) value of $\vert \Delta V_{\text{1 loop (0T)}} \vert$ and benchmark II to have a weak phase transition but relatively tuned vacuum energy difference. For each benchmark we give the main parameter values (see Table~\ref{bench_paraSpec}) and the Higgs mass spectrum (see Table~\ref{bench_massSpec}). The full set of defining parameters is given in Appendix~\ref{section:benchGNMSSM}.

\begin{figure}[t]
\centering
\begin{minipage}[b]{7.5cm}
\centering
\subcaptionbox{\small{Potential at zero temperature.}}{\includegraphics[width=7.5cm]{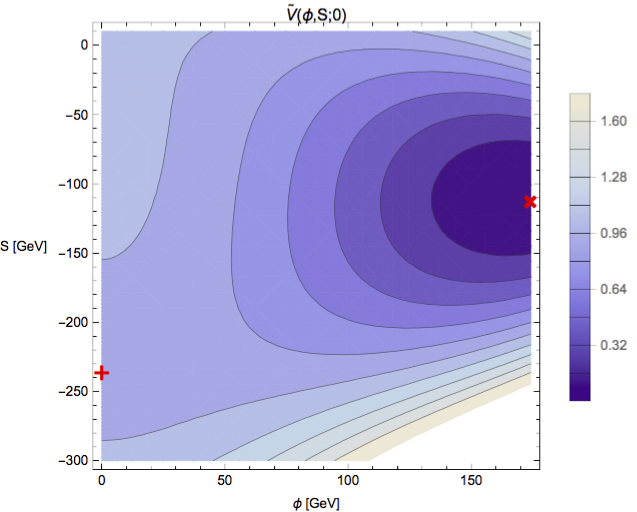}}
\end{minipage}
\begin{minipage}[b]{7.5cm}
\centering
\subcaptionbox{\small{Potential at critical temperature.}}{\includegraphics[width=7.5cm]{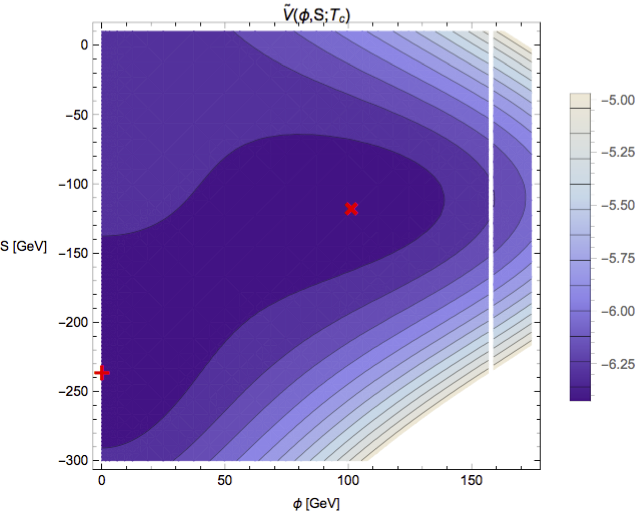}}
\end{minipage}
\caption{\small{The above plots show the shape of the one loop effective potential in $(\phi,S)$ field space at (a) zero temperature and (b) critical temperature for benchmark I. The broken (symmetric) vacuum is marked by a red cross (plus). At zero temperature, the broken and symmetric vacua are located at ($174.2$, $-110.1$) and ($0$,$-234.6$), respectively. At the critical temperature, $T_c=142.5~\text{GeV}$, the broken and symmetric vacua are located at ($101.5$, $-115.4$) and ($0$,$-234.6$), respectively. All fields are in units of GeV. The potential displayed is defined in eq.~(\ref{defV}).}}
\label{benchI}
\end{figure}

\begin{table}[b]
\centering
\small
\begin{tabular}{c | c c c c c c c c c }
\hline
\\
\textbf{Benchmark} & $\lambda$ & $\lambda A_\lambda$ & $k_3$ & $v_S$ & $\tilde{v}_S$ & $m_{\tilde{t}_2}$ & $\Delta V_{\text{1 loop (0T)}}$ & $T_c$ & $\xi$\\
\\
\hline
\textbf{I} & 0.577 & 641.1 & -0.151 & -110.1 & -234.6 & 613.1 & -$1.15\times 10^{8}$ & $142.5$ & 1.01 \\
\textbf{II} & 0.569 & 130.4 & 0.280 & -161.5 & 0.0 & 844.1 & -$6.99\times 10^{7}$ & $116.0$ & 0.49 \\
\textbf{III} & 0.626 & 265.2 & -0.251 & -146.7 & -348.3 & 907.7 & -$6.79\times 10^{6}$ & $47.1$ & 5.20 \\
\hline
\end{tabular}
\caption{\small{Some of the more important quantities for each benchmark scenarios. The full set of parameter values are provided in Appendix~\ref{section:benchGNMSSM}. All masses are in units of GeV.}}
\label{bench_paraSpec}
\end{table}

\begin{table}[b]
\centering
\begin{tabular}{c | c c c c c c c c}
\hline
\\
\textbf{Benchmark} & $\vert\sin\alpha\vert$ & $m_h$ & $m_H$ & $m_s$ & $m_A$ & $m_{A_s}$ & $m_{H^\pm}$\\
\\
\hline
\textbf{I} & 0.119 & 125.0 & 853.7 & 107.2 & 779.1 & 945.0 & 839.1\\
\textbf{II} & 0.013 & 125.0 & 888.0 & 33.0 & 887.1 & 833.3 & 883.5\\
\textbf{III} & 0.172 & 125.0 & 2586.8 & 89.3 & 2586.5 & 1212.0 & 2585.0 \\
\hline
\end{tabular}
\caption{\small{One loop zero temperature Higgs mass spectrum in the benchmark scenarios. All masses are in units of GeV.}}
\label{bench_massSpec}
\end{table}

\begin{table}[b]
\centering
\small
\begin{tabular}{c | c c c}
\hline
\\
\textbf{Benchmark} & $\delta(\phi)$ & $\delta(S)$ & \text{Behaviour}\\
\\
\hline
\textbf{I} & $0.42$ & $0.030$ & \pbox[t]{12.0cm}{\tiny{Minimally strong phase transition, minimal tuning of $\vert \Delta V_{\text{1 loop (0T)}}\vert$}} \\
\textbf{II} & $0.77$ & $0.90$ & \pbox[t]{12.0cm}{\tiny{Weak phase transition, irrespective of the tuning of $\vert \Delta V_{\text{1 loop (0T)}}\vert$}} \\
\textbf{III} & $6.0\times 10^{-5}$ & $1.1 \times 10^{-3}$ & \pbox[t]{12.0cm}{\tiny{Very strong phase transition, significant tuning of $\vert \Delta V_{\text{1 loop (0T)}}\vert$}} \\
\hline
\end{tabular}
\caption{\small{Fractional change of the $\phi$ and $S$ fields using eq.~(\ref{perChange}) and the behaviour of each benchmark.}}
\label{bench_beh}
\end{table}

\begin{figure}[t]
\centering
\begin{minipage}[b]{7.5cm}
\centering
\subcaptionbox{\small{Potential at zero temperature.}}{\includegraphics[width=7.5cm]{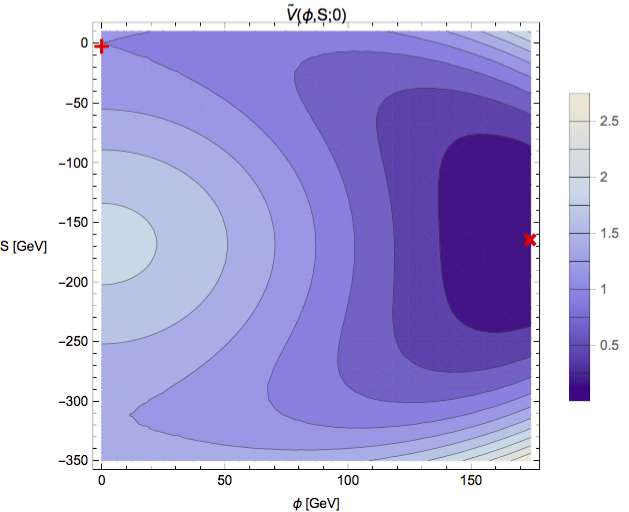}}
\end{minipage}
\begin{minipage}[b]{7.5cm}
\centering
\subcaptionbox{\small{Potential at critical temperature.}}{\includegraphics[width=7.5cm]{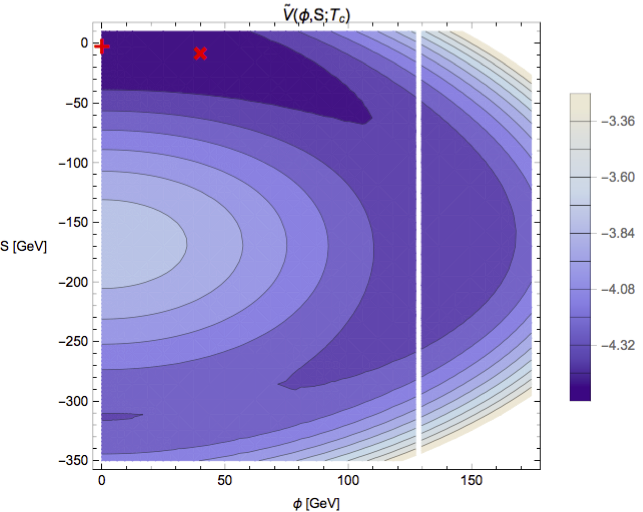}}
\end{minipage}
\caption{\small{The above plots show the shape of the one loop effective potential in $(\phi,S)$ field space at (a) zero temperature and (b) critical temperature for benchmark II. The broken (symmetric) vacuum is marked by a red cross (plus). At zero temperature, the broken and symmetric vacua are located at ($174.2$, $-161.5$) and ($0$,$0$), respectively. At the critical temperature, $T_c=116.0~\text{GeV}$, the broken and symmetric vacua are located at ($40.0$, $-5.19$) and ($0$,$0$), respectively. All fields are in units of GeV. The potential displayed is defined in eq.~(\ref{defV}).}}
\label{benchII}
\end{figure}

\begin{figure}[t]
\centering
\begin{minipage}[b]{7.5cm}
\centering
\subcaptionbox{\small{Potential at zero temperature.}}{\includegraphics[width=7.5cm]{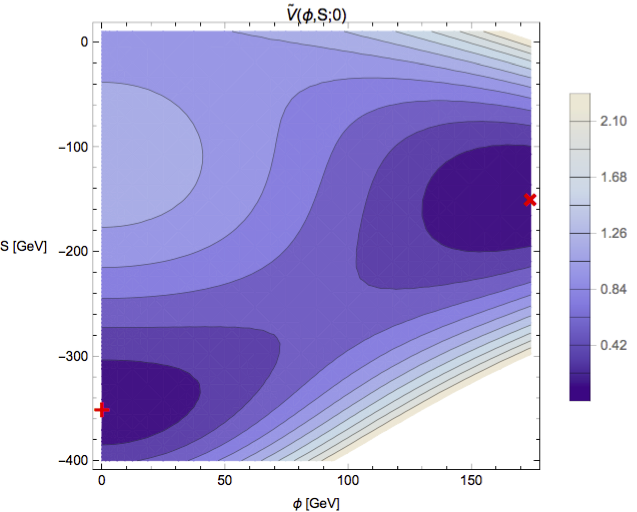}}
\end{minipage}
\begin{minipage}[b]{7.5cm}
\centering
\subcaptionbox{\small{Potential at critical temperature.}}{\includegraphics[width=7.5cm]{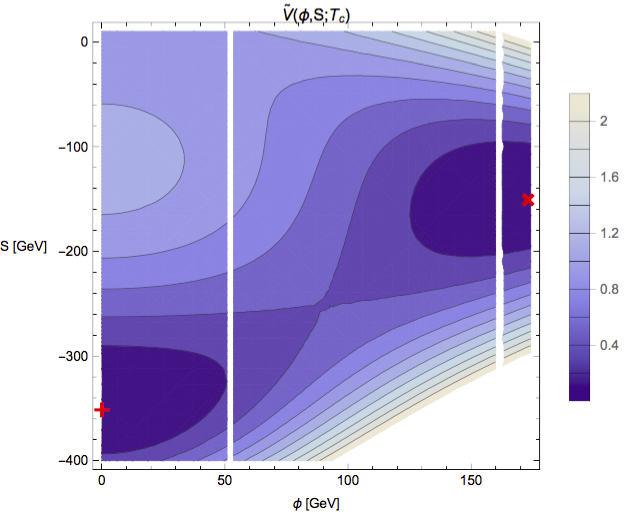}}
\end{minipage}
\caption{\small{The above plots show the shape of the one loop effective potential in $(\phi,S)$ field space at (a) zero temperature and (b) critical temperature for benchmark III. The broken (symmetric) vacuum is marked by a red cross (plus). At zero temperature, the broken and symmetric vacua are located at ($174.2$, $-146.7$) and ($0$,$-348.3$), respectively. At the critical temperature, $T_c=47.1~\text{GeV}$, the broken and symmetric vacua are located at ($173.2$, $-146.9$) and ($0$,$-348.3$), respectively. All fields are in units of GeV. The potential displayed is defined in eq.~(\ref{defV}).}}
\label{benchIII}
\end{figure}

For each benchmark, contour plots of the potential at zero temperature and critical temperature are given in Figures~\ref{benchI}-\ref{benchIII}. The potential displayed in the contour plots is offset and normalised according to
\begin{equation}
\tilde{V}(\phi,S;T)=\dfrac{V_{\text{1 loop}}(\phi,S;T)-V_{\text{1 loop (0T)}}(v,v_S)}{V_{\text{1 loop (0T)}}(0,0)-V_{\text{1 loop (0T)}}(v,v_S)}.
\label{defV}
\end{equation}
Thus the potential in the broken vacuum at zero temperature corresponds to zero in the displayed potential, $\tilde{V}(v,v_S;0)=0$, and the zero-field value of the one loop zero temperature potential corresponds to unity, $\tilde{V}(0,0;0)=1$. The broken (symmetric) vacuum is marked on each potential as a red cross (plus).

The key observation is to see how much the broken vacuum has moved away from its zero temperature value at the critical temperature. Specifically, the singlet value in the broken vacuum does not change by much in benchmarks with a strong phase transition, whereas the singlet value of the broken vacuum changes significantly in benchmark II. To quantify the change of any field value in the broken vacuum, we define the fractional change to be
\begin{equation}
\delta(\Phi)=\dfrac{\vert\Phi_{\text{broken}}(T=0)-\Phi_{\text{broken}}(T=T_c)\vert}{v} \text{,}
\label{perChange}
\end{equation}
where $\Phi$ is to be recognised with one of our fields. A lower fraction corresponds to the field at critical temperature remaining close to its zero temperature value, whereas a high fraction corresponds to the field at critical temperature being far from its zero temperature value. In Table~\ref{bench_beh} we display the values for each benchmark. This allows us to qualitatively link our hypothesis to each of the benchmarks. Namely, that the broken minimum should remain in a neighbourhood of its zero temperature value if we want a strong phase transition.

All of our benchmarks have small Higgs-singlet mixing in accordance with experimental constraints. The singlet state is always lighter than the SM-like Higgs and for benchmark II it is significantly lighter. For all benchmarks the Higgs-singlet coupling $\lambda$ is close to the upper bound that prevents running into a Landau pole \cite{Ellwanger:1996}. All other Higgs states are heavy and decouple from the phase transition.

Benchmark I has a very moderate tuning of $\vert\Delta V_{\text{1 loop (0T)}}\vert$ and does not suffer from a sliding singlet instability, so we arrive at a phase transition with $\xi=1.01$. This is just strong enough to avoid baryon number washout. In Figure~\ref{benchI}, we see that the symmetric and broken minima are well separated by a barrier which does not disappear as we approach the critical temperature, $T_c=142.5~\text{GeV}$. Since the critical temperature is relatively high, the critical Higgs field $v_c$ is noticeably different from its zero temperature value $v$. However, we notice that the singlet hardly moves during the phase transition.

In benchmark III we significantly tune the vacuum energy difference to a small value, whilst keeping the singlet relatively heavy. This results in a very strong first order phase transition with $\xi=5.20$ and a much reduced critical temperature of $T_c=47.1~\text{GeV}$. In Figure~\ref{benchIII} we see a greatly enhanced barrier compared to Figure~\ref{benchI}. Both fields hardly move in this case. We expect the symmetric vacuum to be metastable in this case so the phase transition may not actually take place. This could be checked by computing the energy of the critical bubble which, however, goes beyond the scope of this paper. Starting from this benchmark and reducing the tuning of the vacuum energy difference, we would expect to retain a strong phase transition but enter a regime where the phase transition actually takes place.

Benchmark II is very much different to the already discussed benchmarks, as is apparent in Figure~\ref{benchII}, which contains a valley connecting the symmetric and broken minima. In this case the singlet is rather light. As discussed in the non-supersymmetric case, as the temperature is increased the Higgs mass squared matrix develops a negative eigenvalue and the field slides toward the symmetric minimum. This is indicated by a big change in the singlet field (see Table~\ref{bench_beh}). As a result the critical temperature, $T_c=116.0~\text{GeV}$, is not as low as the vacuum energy difference suggests.

Overall, these benchmarks indicate that a strong first order phase transition can be enforced by having a not too light singlet state with small mixing to the Higgs and a moderately tuned vacuum energy difference.

\section{\label{section:conc}Conclusion}
In this work, we have investigated in detail the one loop vacuum energy difference at zero temperature, $\Delta V_{\text{1 loop (0T)}}$, and its implications on the strength of the electroweak phase transition, $\xi=\sqrt{2}v_c/T_c$. The study was conducted using three single field modifications to the SM, one non-supersymmetric singlet extension to the SM, and a supersymmetric singlet extension (the GNMSSM).

For the single field models investigated, we find that a decrease in $\vert\Delta V_{\text{1 loop (0T)}}\vert$ also decreases the critical temperature. In turn the critical field value remains close to its zero temperature value. This leads to a strong $\xi$-$\Delta V_{\text{1 loop (0T)}}$ correlation with universal behaviour observed at very low $\vert\Delta V_{\text{1 loop (0T)}}\vert$, as can be seen in Figure~\ref{singleField_PtPlot_fig}. This universal behaviour is found in Section~\ref{section:numSingle} to be fixed by the number of relevant degrees of freedom in the plasma. Parameter points with a strong phase transition are guaranteed with only a moderate tuning of the vacuum energy difference, see eq.~(\ref{singleBound}), relative to the SM value in eq.~(\ref{SMvac}).

{
To comment on the reliability of the perturbative techniques used for our analysis, we note that a recent lattice study on the dimension-six extended SM model is found to be consist with the results from the perturbative approach \cite{Akerlund:2015}. They also remark on the observation that it is the Higgs potential itself that determines the nature of the phase transition and not so much the gauge or fermionic degrees of freedom entering through radiative corrections. This supports the idea that higher order loop effects are not crucial in deciding the nature of the phase transition - at least for parameter points that have a strong correlation between $\xi$ and $\Delta V_{\text{1 loop (0T)}}$.}

For singlet extended models, we find a similar $\xi$-$\Delta V_{\text{1 loop (0T)}}$ correlation so long as the fields in the broken vacuum do not slide under thermal effects. This sliding behaviour is most obvious in Section~\ref{section:numNonSUSY} when we look at the non-supersymmetric model with a $\mathbb{Z}_2$ symmetry imposed on the singlet, called the $\mathbb{Z}_2$xSM. We find that a spontaneous breaking of the $\mathbb{Z}_2$ before the critical temperature disfavours a strong phase transition. Such parameter points fall onto an undesirable region in $\xi$-$\Delta V_{\text{1 loop (0T)}}$ space. With the exception of the $\mathbb{Z}_2$xSM unbroken at zero temperature, parameter points on this undesirable region almost disappear completely after imposing phenomenological constraints. This can be seen in Figures~\ref{singleField_PtPlot_fig}, \ref{Z2xSM_PtPlot_fig}, \ref{xSM_PtPlot_fig}, and \ref{GNMSSM_PtPlot_fig}. The reason so many points are removed is because the phenomenological constraints disallow light singlet states with large Higgs-singlet mixing, see Figure~\ref{GNMSSM_PtPlot_mixing_fig}. In other words, phenomenological constraints work in favour of a strong $\xi$-$\Delta V_{\text{1 loop (0T)}}$ correlation.

For the non-supersymmetric singlet extended model with the $\mathbb{Z}_2$ explicitly broken at zero temperature, phenomenological constraints remove the majority of parameter points in our data set. Nonetheless, the surviving points follow the usual $\xi$-$\Delta V_{\text{1 loop (0T)}}$ correlation and a strong phase transition is guaranteed if $\vert \Delta V_{\text{1 loop (0T)}} \vert<1.03 \times 10^8~\text{GeV}^4$.

For the GNMSSM, similar observations to those in the non-supersymmetric singlet extension are made. Three benchmark scenarios are analysed in detail in Section~\ref{section:bench}. Once phenomenological constraints are applied, a strong phase transition is guaranteed if $\vert \Delta V_{\text{1 loop (0T)}} \vert<6.98 \times 10^7~\text{GeV}^4$. However, this is at the cost of excluding a significant portion of the parameter space with a strong phase transition. Instead a far more useful bound is provided in eq.~(\ref{recipepheno}). From Figure~\ref{GNMSSMms0SFOpheno_fig} we can see that this bound captures far more of the parameter space with a strong phase transition.

We stress that this work does not address the  surface tension, tunnelling rate, or the latent heat of the phase transition as measures of the strength of the phase transition. These quantities will indeed depend on the actual height of the barrier, so that we do not expect a universal behaviour correlated to the vacuum energy.

We hope that our results make phenomenological studies with parameters exhibiting a strong phase transition far easier to address. This can be useful for model builders that want a strong phase transition, without the need for any finite temperature calculations.

\begin{center}
\textbf{Acknowledgements}
\\
We would like to thank Thomas Konstandin for useful discussions. C.P.D.H is
supported by the Science and Technology Facilities Council. S.J.H is supported by the
Science and Technology Facilities Council (STFC) under grant ST/J000477/1.
\end{center}

\newpage
\appendix
\appendixpage
\addappheadtotoc

\section{\label{section:paraxSM}Parameter space scan (xSM)}
\noindent Throughout the numerical scan, the ($\mathbb{Z}_2$)xSM parameters are assigned random values following the below table. These parameters are chosen through linear distributions.
\begin{table}[h!]
\centering
\begin{tabular}{c c c c c}
\hline
Parameter: & Mass dimension, $n$: & Minimum: & Maximum: & Determined: \\
\hline
$\lambda_0$ & 0 & $m_{\phi_1}^2/4v^2$ & $m_{\phi_2}^2/4v^2$ & Random assignment\\
$v_S$ & 0 & $-M$ & $0$ & Random assignment\\
$\vert a_1\vert$ & 1 & $0$ & $M$ & Random assignment\\
$\vert b_3\vert$ & 1 & $0$ & $M$ & Random assignment\\
$m_{s}$ & 1 & $0$ & $M$ & Random assignment\\
\hline
$\vert a_2 \vert$ & 0 & $0$ & $10$ & Reparameterisation\\
$b_4$ & 0 & $0$ & $10$ & Reparameterisation\\
$\mu$ & 1 & $-$ & $-$ & Minimum condition\\
$b_2$ & 2 & $-$ & $-$ & Minimum condition\\
$m_{h}$ & 1 & $125$ & $125$ & Fixed\\
\hline
\end{tabular}
\caption{\small{Table of real values randomly assigned to each ($\mathbb{Z}_2$)xSM parameter throughout the numerical scan. The dimension column is given in units of mass dimension, i.e. $[M]^n$. The final column labels how the numerical value is determined.}}
\label{xSM_para_choice}
\end{table}

\newpage
\section{\label{section:paraGNMSSM}Parameter space scan (GNMSSM)}
\noindent Throughout the numerical scan, the GNMSSM parameters are assigned with a natural description for the GNMSSM at low energy scale. This implies that the GNMSSM may be easily described through a top-down approach with a low enough supersymmetry breaking scale, usually $\lesssim\mathcal{O(\text{1 TeV})}$, so as to not demand a huge fine-tuning of the parameters. All parameters are randomly chosen through linear distributions, except for $\vert \lambda \vert$ and $\vert k_3 \vert$ which are determined through log distributions.

\begin{table}[h!]
\centering
\begin{tabular}{c c c c c}
\hline
Parameter: & Mass dimension, $n$: & Minimum: & Maximum: & Determined: \\
\hline
$\tan\beta$ & 0 & $1$ & $10$ & Random assignment\\
$\vert\lambda\vert$ & 0 & $1.0\times 10^{-3}$ & $0.7$ & Random assignment\\
$v_S$ & 1 & $-250$ & $0$ & Random assignment\\
$\vert\mu\vert$ & 1 & $0$ & $m_{\text{SUSY}}$ & Random assignment\\
$\vert k_1\vert$ & 2 & $0$ & $m_{\text{SUSY}}^2$ & Random assignment\\
$\vert k_2\vert$ & 1 & $0$ & $m_{\text{SUSY}}$ & Random assignment\\
$\vert k_3\vert$ & 0 & $1.0\times 10^{-3}$ & $0.7$ & Random assignment\\
$\vert b \mu \vert$ & 1 & $0$ & $m_{\text{SUSY}}^2$ & Random assignment\\
$\vert k_2 A_{k_2} \vert$ & 1 & $0$ & $m_{\text{SUSY}}^2$ & Random assignment\\
$\vert k_3 A_{k_3}\vert$ & 1 & $0$ & $m_{\text{SUSY}}$ & Random assignment\\
$\vert \lambda A_\lambda\vert$ & 1 & $0$ & $m_{\text{SUSY}}$ & Random assignment$~^*$\\
\hline
$m_{Q_3}$ & 1 & $\Delta m_3$ & $m_{\text{SUSY}}$ & Fixed for $m_h$\\
$m_{\bar{u}_3}$ & 1 & $-$ & $-$ & Fixed\\
$\Delta m_3$ & 1 & $100$ & $100$ & Fixed\\
$A_t$ & 1 & $-$ & $-$ & Fixed\\
$m_{h^0}$ & 1 & $125$ & $125$ & Fixed\\
$m_{H_u}$ & 1 & $-$ & $-$ & Minimum condition\\
$m_{H_d}$ & 1 & $-$ & $-$ & Minimum condition\\
$m_{S}$ & 1 & $-$ & $-$ & Minimum condition\\
$\vert A_{k_1}\vert$ & 1 & $0$ & $m_{\text{SUSY}}$ & No linear term in $S$\\
\hline
\end{tabular}
\caption{\small{Table of real values randomly assigned to each GNMSSM parameter throughout the numerical scan. The dimension column is given in units of mass dimension, i.e. $[M]^n$. The final column labels how the numerical value is determined.}
\newline
\small{$~^*$Note that the $A_\lambda$ parameter is randomly assigned subject to the broken vacuum being the absolute minimum of the potential.}}
\label{rand_para_choice}
\end{table}

\newpage
\section{\label{section:benchGNMSSM}GNMSSM benchmarks: parameter points}
\noindent The assigned parameter values for each of the benchmark scenarios is provided in the table below.

\begin{table}[h!]
\centering
\begin{tabular}{c c c c c}
\hline
Parameter: & Benchmark I: & Benchmark II: & Benchmark III: \\
\hline
$\tan\beta$ & $1.350$ & $2.355$ & $5.133$\\
$\lambda$ & $0.5770$ & $0.5690$ & $0.6266$\\
$v_S~\text{[GeV]}$ & $-110.1$ & $-161.5$ & $-146.7$\\
$\mu~\text{[GeV]}$ & $463.7$ & $275.5$ & $278.6$\\
$k_1~\text{[GeV]}^2$ & $-6.820\times 10^{5}$ & $-7.547 \times 10^{5}$ & $8.624\times 10^{5}$\\
$k_2~\text{[GeV]}$ & $-303.7$ & $367.8$ & $529.2$\\
$k_3$ & $-0.1513$ & $0.2804$ & $-0.2508$\\
$b \mu~\text{[GeV]}^2$ & $7.843\times 10^{5}$ & $7.621\times 10^{5}$ & $8.057\times 10^{5}$\\
$k_2 A_{k_2}~\text{[GeV]}^2$ & $-6.072\times 10^{5}$ & $3.440\times 10^{4}$ & $-2.065\times 10^{5}$\\
$k_3 A_{k_3}~\text{[GeV]}$ & $-124.5$ & $-233.8$ & $456.6$\\
$\lambda A_\lambda~\text{[GeV]}$ & $641.1$ & $130.4$ & $265.2$\\
$m_{Q_3}~\text{[GeV]}$ & $688.8$ & $926.7$ & $991.7$\\
\hline
\end{tabular}
\caption{\small{Table of values assigned to each of the considered GNMSSM benchmark scenarios.}}
\label{bench_para}
\end{table}

\end{document}